\documentclass[superscriptaddress,preprint,showkeys,aip,cha]{revtex4-1}

\usepackage[dvips]{graphicx}
\usepackage[utf8]{inputenc}
\usepackage{multirow,booktabs}
\usepackage{amsmath,amssymb,latexsym,MnSymbol}

\usepackage[table]{xcolor}
\usepackage{xcolor}
\usepackage[%
  colorlinks=true,
  urlcolor=blue,
  linkcolor=blue,
  citecolor=blue
]{hyperref}

\begin{document}
\title{The hidden traits of endemic illiteracy in cities}

\author{Luiz G. A. Alves}
\affiliation{Institute of Mathematics and Computer Science, University of S\~ao Paulo, S\~ao Carlos, SP 13566-590, Brazil}

\author{Jos\'e S. Andrade Jr.}
\affiliation{Departamento de F\'isica, Universidade Federal do Cear\'a, Fortaleza, CE, 60451-970, Brazil}

\author{Quentin S. Hanley}
\affiliation{School of Science and Technology, Nottingham Trent University, Clifton Lane, Nottingham NG11 8NS, United Kingdom}

\author{Haroldo V. Ribeiro}\email{hvr@dfi.uem.br}
\affiliation{Departamento de F\'isica, Universidade Estadual de Maring\'a, Maring\'a, PR 87020-900, Brazil}

\begin{abstract}
In spite of the considerable progress towards reducing illiteracy rates, many countries, including developed ones, have encountered difficulty achieving further reduction in these rates. This is worrying because illiteracy has been related to numerous health, social, and economic problems. Here, we show that the spatial patterns of illiteracy in urban systems have several features analogous to the spread of diseases such as dengue and obesity. Our results reveal that illiteracy rates are spatially long-range correlated, displaying non-trivial clustering structures characterized by percolation-like transitions and fractality. These patterns can be described in the context of percolation theory of long-range correlated systems at criticality. Together, these results provide evidence that the illiteracy incidence can be related to a transmissible process, in which the lack of access to minimal education propagates in a population in a similar fashion to endemic diseases.
\end{abstract}
\maketitle
\section*{Introduction}

The world has experienced unprecedented progress towards eradicating illiteracy since the mid-twentieth century. According to UNESCO, the illiteracy rate at world level has decreased from $44.3\%$ in the 50s to about $14\%$ in 2015~\cite{UNESCO}. While this progress is impressive, the number of illiterate people has increased from 700 to 745 million over the same period because of the rapid population growth. The reduction of illiteracy is not equally distributed over the globe and disproportionately affects women. There exist countries where illiteracy has remained stubbornly high, such as in Sub-Saharan Africa, and Oceania has seen illiteracy rates increase~\cite{UNESCO}. Even developed countries have encountered notable difficulties to continue reducing illiteracy rates. For instance, the latest available study carried out by the US Department of Education found no significant change in the illiteracy rate among adults between 1992 and 2003~\cite{kutner2006first}, which was estimated to be around 14\%. This scenario is quite worrying because illiteracy has been associated with health problems~\cite{dewalt2004literacy,berkman2011low} such as diabetes~\cite{schillinger2002association}, hypertension~\cite{williams1998relationship}, depression~\cite{gazmararian2000multivariate}, and schizophrenia~\cite{liu2013illiteracy}. It is also related to unhealthy habits such as smoking~\cite{gavarasana1992illiteracy}, violent behavior~\cite{davis1999low,reportpolice}, and reduced life expectancy~\cite{messias2003income}. While the precise economic costs worldwide are difficult to quantify, estimated annual losses due to illiteracy  are in the billions of dollars in the US alone~\cite{baker1997relationship,roman2004illiteracy}, resulting mainly from health-related care costs, low productivity, and strains on the welfare system. 

This survey of the literature makes clear that illiteracy poses devastating effects on individuals, the economy, and society in general. Thus, it is essential to understand the underlying mechanisms that have hampered the reduction in illiteracy rates over the world. Illiteracy has been long recognized as an inter-generational trend~\cite{costa1988adult,roman2004illiteracy}, that is, similar to genetic disorders, illiteracy may be passed on from parent to child. Other studies~\cite{alves2013distance,alves2015scale} have shown that cities with high illiteracy rates exhibit poor performance in reducing illiteracy in the future, whereas cities with low rates tend to display even lower illiteracy rates in future. These studies suggest that illiteracy propagates through family and social networks. This idea is supported by the recent works of Christakis and Fowler~\cite{christakis2009connected}, which have demonstrated that individual features such as obesity~\cite{christakis2007spread}, smoking habits~\cite{christakis2008collective}, and happiness~\cite{fowler2008dynamic} spread through networks in a population in a manner similar to infectious diseases. Although it has not been empirically verified yet, the hypothesis that illiteracy behaves like traditional transmissible
diseases naturally emerges within this context. The paucity of studies addressing this issue reflects the enormous challenge of following an empirical social network (containing a few thousand people) during enough time to observe the possible spreading dynamics of illiteracy. Even though the works of Christakis and Fowler demonstrate the feasibility of such approaches for some individual features, such datasets are still quite rare.

To overcome this shortage of detailed data and test the hypothesis that illiteracy exhibits characteristics of transmissible diseases, we investigated spatial patterns in the incidence of illiteracy over a system of cities. Our approach is motivated by the fact that infectious diseases spreading through urban systems show long range correlations, cluster formation, and fractality~\cite{kulldorff1995spatial,tilman1997spatial,grenfell2002dynamics,viboud2006synchrony,riley2007large,chowell20081918,pitzer2009demographic,balcan2009multiscale,gallos2012collective,viboud2013contrasting,brockmann2013hidden,gog2014spatial,sun2016pattern,antonio2017spatial} . By probing these spatial fingerprints in patterns of illiteracy and comparing with those exhibited by infectious diseases, we should be able to uncover supporting evidence for the hypothesis of an epidemic-like spreading of illiteracy. 

By using data from all Brazilian cities (over 5000) in three different years, we show that illiteracy rates are long-range correlated and present a non-trivial cluster structure, characterized by a heavy-tailed distribution and fractal dimensions very close to those reported for diseases. Our results reveal that the spatial patterns of illiteracy incidence in cities are strikingly similar to those observed for infectious diseases providing indirect evidence that illiteracy incidence may be driven by a transmissible process, information that may help in the creation of better public policies and strategies for reducing the prevalence of illiteracy.

\section*{Results and Discussion}

\begin{figure}[!ht]
\centering
\includegraphics[width=0.65\linewidth]{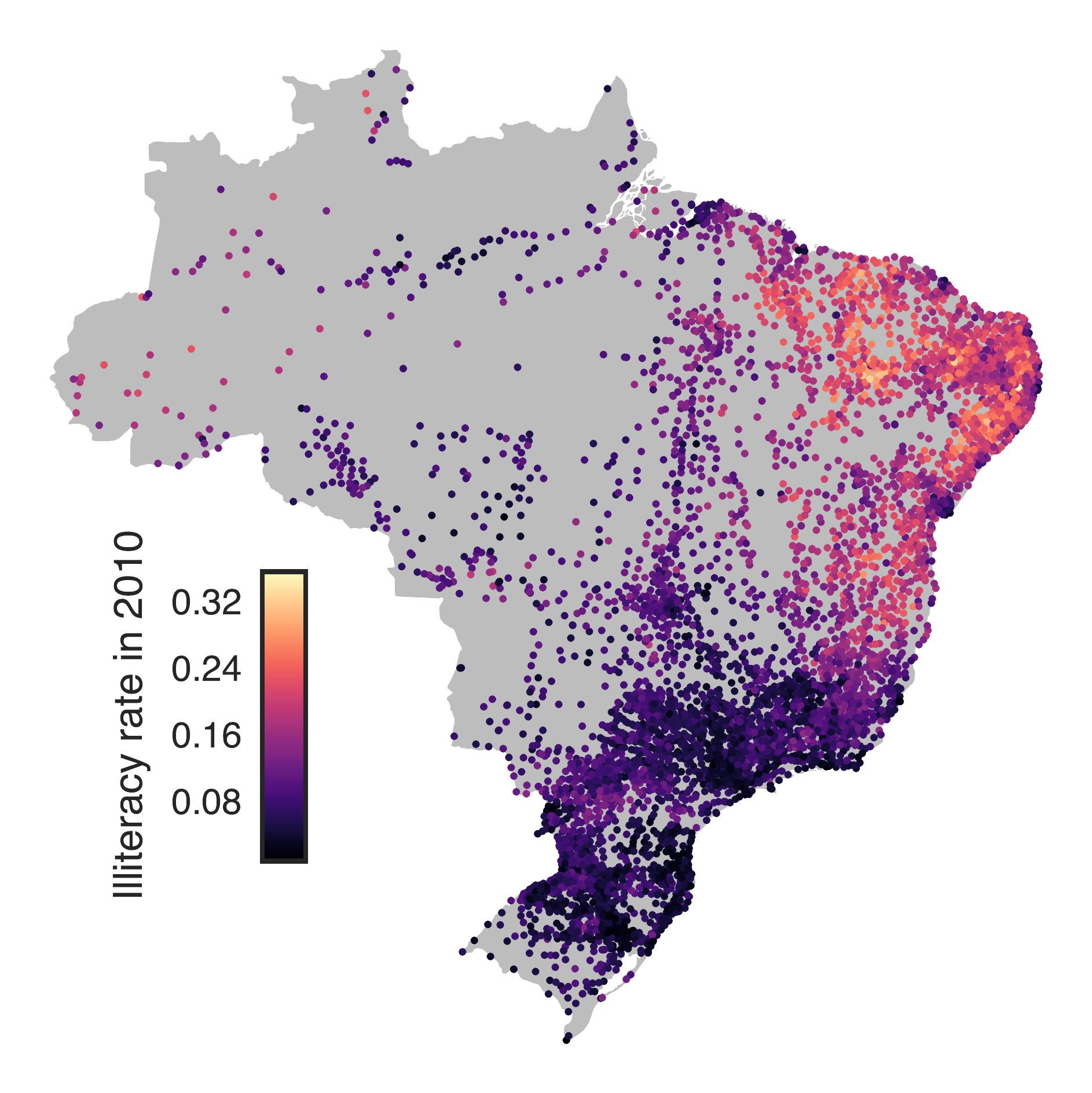}
\caption{{\bf Mapping illiteracy among Brazilian cities.} Each dot on this map represents the location of a Brazilian city and the color code indicates the illiteracy rate at that place for the year 2010. We note that most cities with high rates are located in the Northeastern region, whereas most Southeast/South cities display small illiteracy rates. We further observe that cities with similar illiteracy rates appear to form clusters. These spatial patterns are quite similar to the other two years in our dataset (\hyperref[sfig1]{Figure S1}).}
\label{fig:1}
\end{figure}

The data used in this study is based on the three latest Brazilian census that took place in 1991, 2000, and 2010. It consists of the per capita number of illiterate people (or illiteracy rate) for each Brazilian city in the three previously-mentioned years and the geographic location of each city (see Methods for details). Figure~\ref{fig:1} illustrates this dataset for the latest census year. This map shows that similar to what happens at world level, illiteracy rates are not evenly distributed among the Brazilian municipalities. Illiteracy rates range from less than 1\% to over 30\% and the map exposes a remarkable spatial segregation splitting the country into two parts. In the Northeastern region, there is a concentration of a large number of cities with high illiteracy rates. In contrast, most Southeast/South cities usually display small illiteracy rates. Similar to worldwide trends, illiteracy in Brazil sharply decreased from over 65\% at the beginning of the twentieth century to less than 10\% in 2010~\cite{MEC}. In spite of this sharp decline as a percentage, the absolute number of illiterate people systematically increased between 1900 and 1980  from 6.3 to 19 million people. The illiterate population only started to decrease in the 90s~\cite{MEC} and  minimal progress has been made over the last decade.

\begin{figure*}[!ht]
\centering
\includegraphics[width=0.9\linewidth]{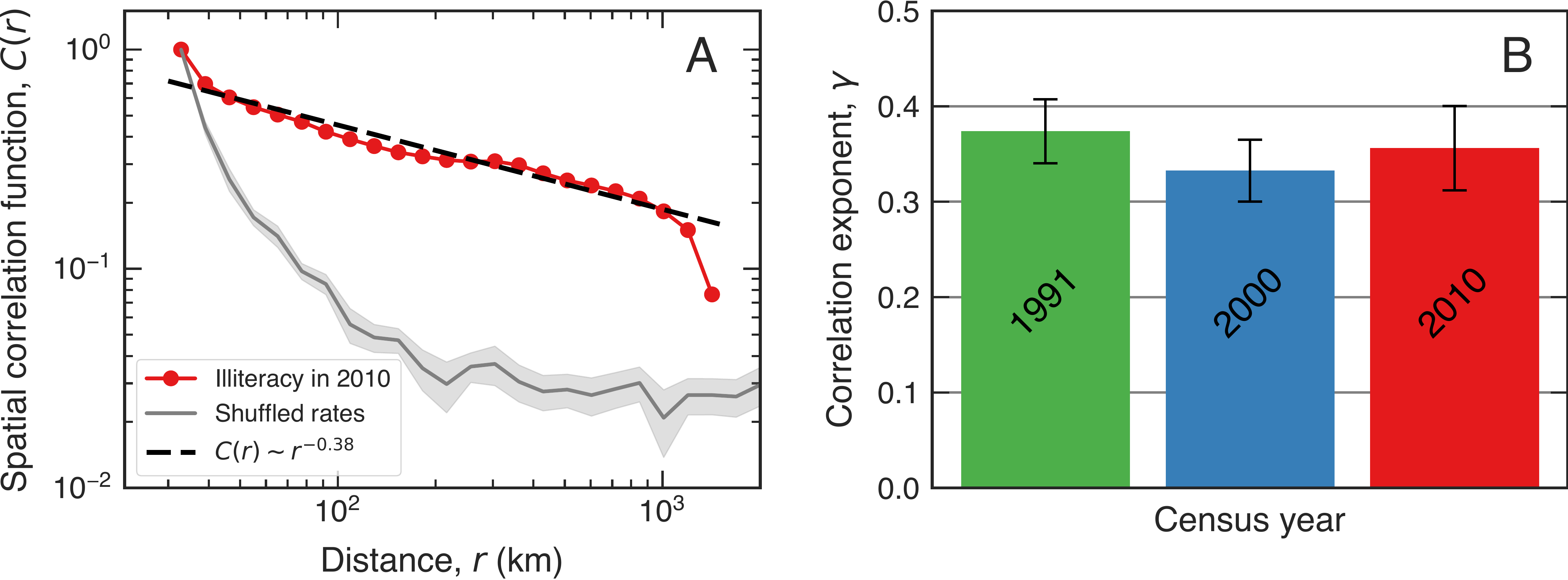}
\caption{{\bf Illiteracy rates are spatially long-range correlated.} (A) The correlation function $C(r)$ of the illiteracy rates in Brazilian cities in the year 2010. The red dots are the empirical values of $C(r)$ and the dashed line is a power-law decaying function, $C(r)\sim r^{-\gamma}$, with $\gamma=0.38$ adjusted to data with $r_{min}=10$~km. The gray curve represents the average values of $C(r)$ after random shuffling the rates among the cities (1000 realizations) and the shaded area stands for the 95\% bootstrap confidence region. Very similar behavior is observed in the other two years (\autoref{sfig2}). (B) Average values of $\gamma$ obtained by least-square fitting the relationship between $\ln C(r)$ and $\ln r$ in the range $r_{min}\leq r\leq1000~\text{km}$, over different values of $r_{min}$ (\autoref{sfig2} for details) and for the three census years. The error bars are 95\% bootstrap confidence intervals.}
\label{fig:2}
\end{figure*}

We start by estimating the spatial correlation function $C(r)$ of the illiteracy rate to quantify the inter-relationships among cities distant by $r$ kilometers (see Methods Section for details). The spatial correlation function measures the average tendency of cities (at a distance $r$) to display similar illiteracy rates (relative to the average rate). A value of $C(r)$ close to 1 implies that rates are strongly correlated, whereas a value close to zero indicates that rates are uncorrelated. \autoref{fig:2}A depicts the behavior of $C(r)$ for the year 2010, where we observe a value close to 1 at short distances and a slow decay of $C(r)$ as $r$ increases. This decay is much slower than the correlation function obtained after random shuffling of the rates among cities (gray curve). For instance, at $r\approx200$~km the correlation function is $\approx0.35$, while after shuffling it is $\approx0.03$. We further observe a cutoff-like behavior for distances greater than $1000$~km, a finite-size effect related to the dimensions of the Brazilian territory. The shape of $C(r)$ is well approximated by a power-law of the form $C(r)\sim r^{-\gamma}$ with $\gamma=0.38$.  Similar behavior is observed for the other two census data (\autoref{sfig2}A). However, it is worth noting that the values of $\gamma$ display small changes depending on the range employed to fit ($r_{min}\leq r\leq1000~\text{km}$), as depicted in~\autoref{sfig2}B. Because of that, we calculate the average value of $\gamma$ over a range values of $r_{min}$ for each year in our dataset. The average values are reported in~\autoref{fig:2}B, where we observe that the exponent $\gamma$ is practically the same for the three census years.

The empirical values of $\gamma$ are much less than 2 which is the value expected for uncorrelated two-dimensional data. These values are also smaller than what is typically observed for population size in US counties~\cite{gallos2012collective} and Brazilian cities~\cite{alves2015spatial} ($\gamma\approx 1$), indicating that population growth alone cannot explain the spatial dynamics of illiteracy. More intriguing, the values of $\gamma$ are between  those reported for obesity and diabetes in the US~\cite{gallos2012collective} ($\gamma\approx0.5$) and dengue in Brazil~\cite{antonio2017spatial} ($\gamma\approx0.3$). Thus, we have confirmed that illiteracy rates among Brazilian cities are long-range correlated in a similar fashion to disease cases in cities. These long-range correlations are found in many physical systems near the criticality such as ferromagnets~\cite{stanley1971introduction} and also in biological systems such as in the brain~\cite{schneidman2006weak,mora2011biological} and bird flocks~\cite{cavagna2010scale}. This behavior is consistent with the hypothesis of a transmissible process of illiteracy. It is still worth remembering that in the case of obesity~\cite{gallos2012collective}, the works of Christakis and Fowler have indeed revealed the epidemic nature of obesity via social network analysis~\cite{christakis2007spread,christakis2009connected}.

In addition to long-range correlations, the emergence of non-trivial cluster structures is another important spatial fingerprint of diseases spreading. To evaluate the presence of such structures in illiteracy rates, we have employed the density-based spatial clustering of applications with noise (DBSCAN)~\cite{ester1996density} algorithm for discovering spatial clusters of cities with similar illiteracy rates. The DBSCAN works by finding core points and them expanding the clusters to points in their neighborhoods. This algorithm has two main parameters: the minimal number of points in a neighborhood for defining a core point ($n_{min}$) and the maximum distance between two points determining a neighborhood ($\varepsilon$). It is worth noting that the DBSCAN is somehow similar to the city clustering algorithm (CCA)~\cite{rozenfeld2008laws}, an approach often used for systematically defining urban units that was also employed for studding obesity clusters in the US~\cite{gallos2012collective}. In our case, we have fixed $n_{min}=1$ (for allowing clusters of unitary size) and explored a range of values for $\varepsilon$ to enhance the universality of our findings. We have investigated the formation of spatial clusters through a percolation-like analysis~\cite{gallos2012collective}, where the DBSCAN algorithm is applied to the set of cities having illiteracy rates larger than $i^*$. Thus, by exploring a range of values for $i^*$, we probe detailed patterns about the formation of clusters at different illiteracy scales and investigate the process from which these clusters grow and merge as the value of $i^*$ decreases.

\begin{figure*}[!ht]
\centering
\includegraphics[width=0.9\linewidth]{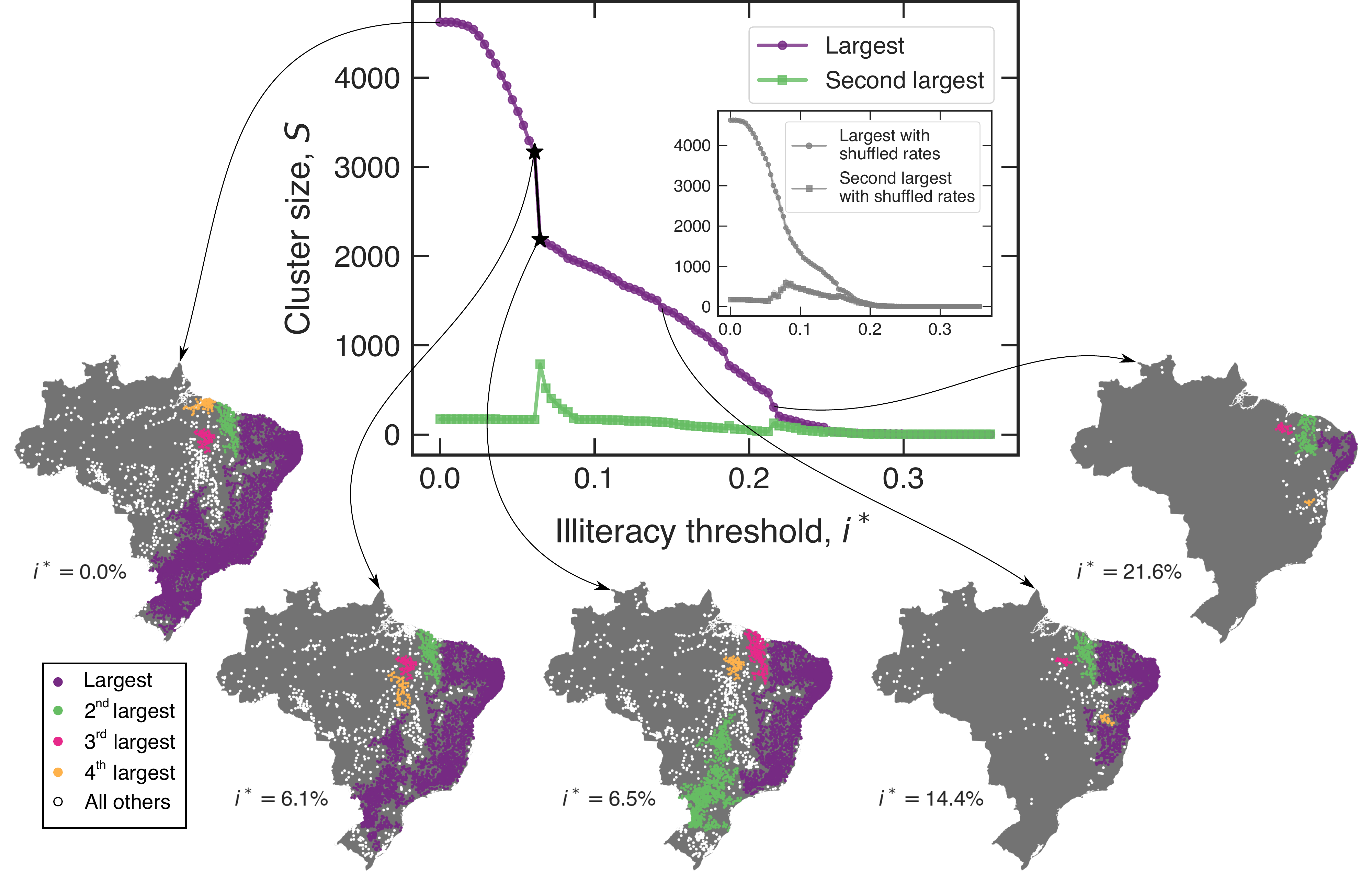}
\caption{{\bf Clustering formation and percolation-like transitions in illiteracy rates.}  The purple curve in the main plot shows the size of the largest cluster of cities ($S$) as a function of the illiteracy threshold ($i^*$). The green curve represents the same for the second largest cluster. We notice a sharp decrease in the value of $S$ and a peak in the size of the second largest component when $i^*\approx 6.1\%$. We further observe other minor sudden changes in these quantities as the value of $i^*$ increases. The maps show the four largest identified clusters (colored in purple, green, pink, and yellow) for particular values of $i^*$ (indicated by the arrows). The white dots represent the cities in other smaller clustering components. See~\autoref{sfig6} for more detailed maps. The inset in the main plot illustrates the behavior of the size of the largest cluster (and second largest) after shuffling the rates among cities. All results are based on 2010 data (see~\autoref{sfig3} for all years) with $\varepsilon=48$~km.}
\label{fig:3}
\end{figure*}

\autoref{fig:3} shows the dependence of the size of the largest (and 2nd largest) cluster on the threshold $i^*$ for the year 2010 and $\varepsilon=48$~km. We notice that the largest cluster encompasses practically all Brazilian cities for $i^*=0$. By increasing $i^*$, the size of the largest cluster decreases. However, differently from uncorrelated percolation~\cite{bunde2012fractals}, where the largest cluster breaks into spatially uniform distributed small clusters, the main cluster of Brazilian cities displays a more complex behavior marked by a sudden change around the value $i^*=6.1\%$. For $i^*$ slightly larger than this threshold, the largest cluster breaks apart into two main components (maps of~\autoref{fig:3}): one including most Northeastern cities and another related to Southern and Midwest cities. These two distinct regions point to the existence of a ``barrier'' separating both groups of cities. Researchers have observed that the Appalachian Mountains may act as a physical barrier for the spreading of obesity among US cities~\cite{gallos2012collective}. However, in our case it is improbable that these two groups of cities are separated by any physical barrier (even of infrastructure origin); instead, this separation is more likely to reflect some ``socioeconomic barrier'' related to the historical formation of the Brazilian cities. Another interesting aspect of this clustering analysis is the peak in the size of the second largest cluster around the value $i^*=6.1\%$, a fingerprint of percolation transitions~\cite{bunde2012fractals}. By continuously increasing the value of $i^*$, we observe a hierarchical process in which these clusters are successively broken into smaller ones (maps of \autoref{fig:3} and \autoref{sfig8}). This process is also marked by other minor sudden changes in the size of the largest cluster and peaks in the size of the second largest component. For very large threshold rates $i^*$, we find that the epicenter of endemic illiteracy is located in the Northeastern region of Brazil. 

Very similar results are obtained for the other two census (\autoref{sfig3}, \autoref{sfig4}, \autoref{sfig5}, \autoref{sfig6}) with $\varepsilon=48$~km. However, we observe that the threshold values of $i^*$ in which the transitions in the size of the largest cluster occur have shifted toward smaller values for more recent years. For instance, in 1991 the transition is observed at $i^*\approx12\%$, whereas it occurs at $i^*\approx 8\%$ and $i^*\approx 6\%$ in the years 2000 and 2010, respectively. On the other hand, the change in the size of the largest cluster has become sharper in the two more recent census. The size of the jump has increased from $\approx400$ in 1991 to $\approx800$ cities in 2000 and 2010. While the decreasing behavior in the threshold values of $i^*$ reflects the overall declining trend of the illiteracy rates (which was more accentuated between 1991 and 2000), the larger jumps in the size of the largest component suggest that the spatial segregation among Brazilian cities has intensified in more recent years. 

It is worth noting that part of these clustering patterns could be related to the spatial distribution of cities. In order to test to which extent the location of Brazilian cities is responsible for the observed results, we have carried out the same percolation-like analysis after randomly shuffling the illiteracy rates among cities. In this way, the long-range correlations among the rates are destroyed and the clustering patterns should be only associated with the spatial distribution of cities. The inset of~\autoref{fig:3} shows the behavior of the size of the largest and 2nd largest cluster as a function of $i^*$ for 2010 data. Also, \autoref{sfig3} depicts these two quantities for all years. For all years, we observe that the sudden decrease in the largest component vanishes when considering the shuffled data; other smaller sudden decreases also disappear after shuffling the illiteracy rates among cities. For the year 1991, we observe that the shuffled results for the second largest component is marked by a peak located at a value of $i^{*}$ very close to the one observed for the actual data; however, other minor peaks in the size of the second largest component are only observed in the actual data. The behavior is slightly different for the years 2000 and 2010. In these cases, we observe that the main peaks in the second largest component emerge at a smaller value of $i^{*}$ when compared with the results obtained from the shuffled data. We further note the staircase-like behavior in the second largest component vanishes after shuffling the rates. Thus, the results obtained when illiteracy rates are randomly shuffled among cities are similar to what is observed in uncorrelated percolation process~\cite{bunde2012fractals}, and therefore, the spatial distribution of cities has a minor role in the clustering results obtained with the actual data.

Naturally, our clustering analysis is affected by the value of the DBSCAN parameter $\varepsilon$, since too small values prevent the formation of clusters, while too large values tend to group all cities~\cite{schubert2017dbscan}. However, the value $\varepsilon=48$~km is arbitrary and our results and conclusions are very robust for $\varepsilon$ between $\approx25$~km and $\approx75$~km (\autoref{sfig7}, \autoref{sfig8}, \autoref{sfig9}, and \autoref{sfig10}). No clustering structures are observed for $\varepsilon<20$~km, whereas for $75>\varepsilon>100$~km, clusters are still formed but the transitions are less-sharp. For even larger values of $\epsilon$, the clustering structure becomes meaningless, and the results are similar to those of an uncorrelated percolation process. 

\begin{figure*}[!ht]
\centering
\includegraphics[width=0.94\linewidth]{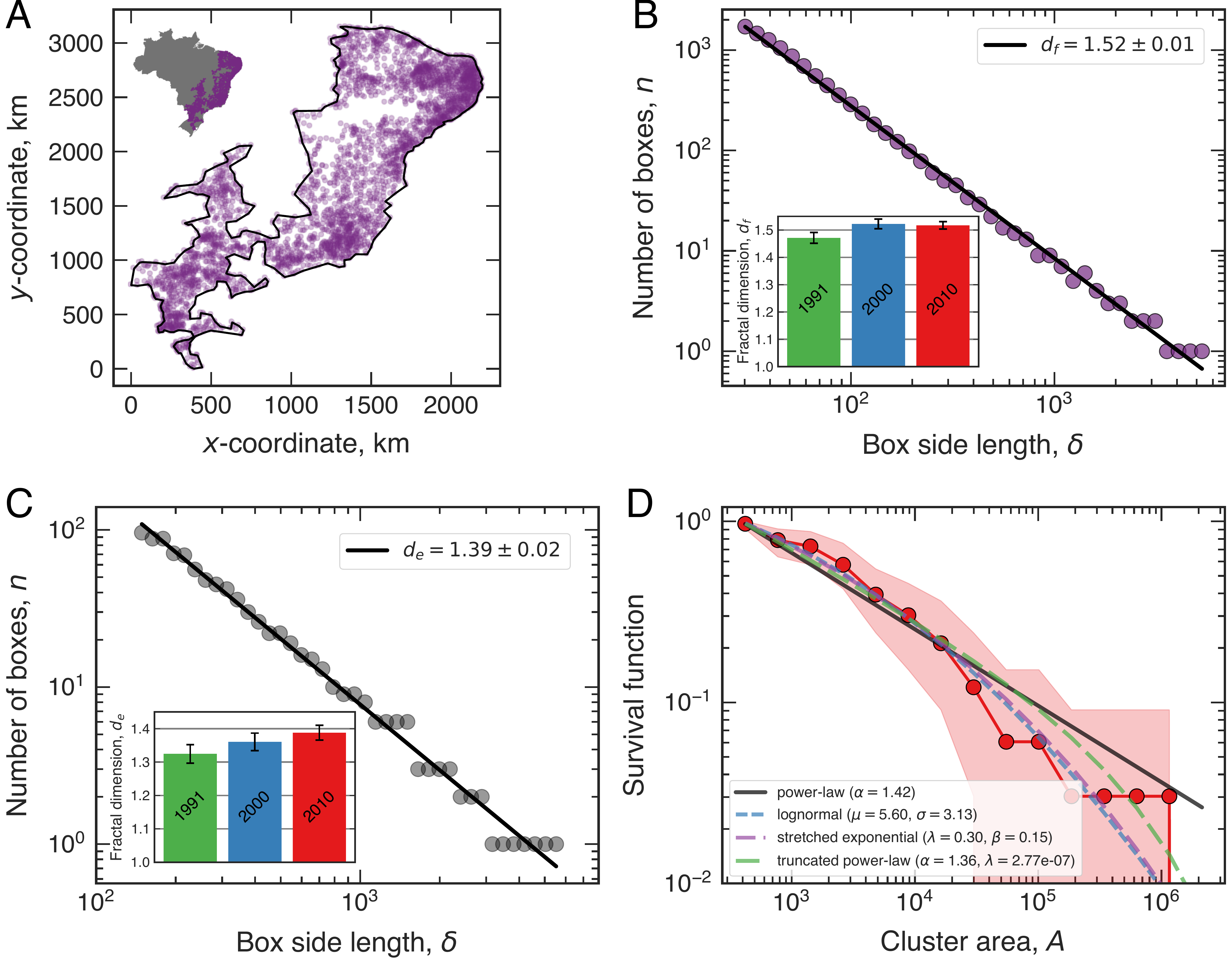}
\caption{{\bf The critical exponents of illiteracy clusters.} (A) The shape of the largest cluster immediately before the transition ($i^*=6.1$) for the year 2010. Each purple dot represents a city within the largest component, and the black line is the concave hull enveloping this cluster. (B) The relationship between the number of boxes $n$ necessary to cover the largest cluster as a function of the box side length $\delta$. The continuous line is a power-law fit ($n\sim\delta^{-d_f}$) with $d_f=1.55\pm0.01$ representing the box-counting fractal dimension (inset shows the values for the other years). (C) The analogous of the previous plot when considering only the concave hull points. The continuous line is a power-law fit ($n\sim\delta^{-d_e}$) with $d_e=1.38\pm0.02$ representing the box-counting fractal dimension of the hull points (inset shows the values for the other years). (D) The survival function (complementary cumulative distribution) of the area of clusters ($A$) near the criticality (red dots). The different lines represent four probability distributions adjusted to the empirical data (names and parameters are shown within the plot legend). The shaded area stands for the 95\% bootstrap confidence region of the survival function. \autoref{sfig11} shows the results for each census year.}
\label{fig:4}
\end{figure*}

Other remarkable spatial properties that have been observed for diseases spreading are the three critical exponents related to clustering formation in long-range correlated systems near the percolation transition~\cite{bunde2012fractals,gallos2012collective}. Two of these exponents are associated with the fractal geometry of the largest cluster: one is the box-counting fractal dimension of the largest cluster ($d_f$), and the other is the fractal dimension of the set points forming the concave hull enveloping the largest cluster ($d_e$). For obesity in the US it was found that $d_f\approx1.8$ and $d_e\approx1.4$~\cite{gallos2012collective}. In our case, \autoref{fig:4}A shows the shape of the largest cluster immediately before the transition in the year 2010~\footnote{We have employed $\varepsilon=48$~km for all fractal analysis, but results are very robust for $25<\varepsilon<75$~km.}. This plot also depicts the concave hull points obtained through a method based on the $k$-nearest neighbors algorithm with $k=1$~\cite{moreira2007concave}. Very similar shapes are obtained for the other two census years (\autoref{sfig11}). \autoref{fig:4}B shows the number of boxes $n$ (of size $\delta\times\delta$) necessary to cover all data points in the largest cluster as a function of $\delta$. The box-counting dimension is defined by fitting the power-law function $n\sim \delta^{-d_f}$ to these data, and yields $d_f=1.52\pm0.01$ (via an ordinary-least-squares fit of the relationship $\log n$ versus $\log \delta$). Similar values are obtained for the other two census (inset of \autoref{fig:4}B and \autoref{sfig11}). \autoref{fig:4}C shows the analogous analysis for the concave hull points, where we have found $d_e=1.38\pm0.02$ for the year 2010. The other two census years have similar values, and a slightly increasing trend is observed (inset of \autoref{fig:4}C and \autoref{sfig11}). The values of $d_e$ are very close to those reported for obesity, while the $d_f$ values are somewhat smaller~\cite{gallos2012collective}, suggesting that the inner spatial structures of the illiteracy cluster are rougher than those of obesity. 

The last exponent is related to the probability distribution of the area of clusters ($A$) near the percolation transition. Percolating systems with long-range correlations usually exhibit a power-law distribution, $P(A)\sim A^{-\alpha}$, where $\alpha$ is the third critical exponent. To calculate this distribution, we have first estimated the concave hull points (also with $k=1$~\cite{moreira2007concave}) of each cluster containing more than three cities and then integrated over these points to evaluate the area $A$. \autoref{fig:4}D shows the empirical form of $P(A)$ for the year of 2010. This distribution clearly has a long-tail, but its curved shape indicates that a power-law is far from being a perfect fit. Even so, we have applied the procedure of Clauset \textit{et al.}~\cite{clauset2009power} for fitting power-law distributions, finding an exponent $\alpha=1.42$ for $A>A_{min}\approx388$~km$^2$ and a $p$-value of the Kolmogorov-Smirnov test equal to $0.18$. While this $p$-value shows that the power-law hypothesis cannot be rejected from data, it does not rule out that other distributions may fit the data better. We have fitted the empirical values of $P(A)$ with lognormal [$P(A)\sim A^{-1} e^{-(\ln A -\mu)^2/(2\sigma^2)}$, $\mu$ and $\sigma$ are the parameters], Weibull [$P(A)\sim A^{\beta-1} e^{-\lambda A^{\beta}}$, $\beta$ and $\lambda$ are the parameters], and truncated power-law [$P(A)\sim A^{-\alpha} e^{-\lambda A}$, $\alpha$ and $\lambda$ are the parameters] distributions. All these fitted distributions are shown in~\autoref{fig:4}D, and despite all Kolmogorov-Smirnov $p$-values being larger than $0.05$, we observe that none of these functions a very good fit to the empirical values of $P(A)$.
We have further compared these three distributions with the power-law via likelihood ratio tests, as shown in~\autoref{table:lr}. The statistical tests show that the lognormal, stretched exponential, and truncated power-law distributions have higher maximum likelihood estimates than the power-law distribution (that is, the log-likelihood ratios are negative). However, the $p$-values of these comparisons indicate that this difference is not statistically significant. Thus, the results of the statistical tests do not allow us to choose among these four probability distributions.

Similar results are obtained for the other two census points (\autoref{sfig11}). For obesity in the US, researchers have found a power-law distribution for the areas of clusters but with $\alpha\approx2$~\cite{gallos2012collective}. Models and simulations describing percolation through nearest neighbors in long-range correlated systems predicts that $\alpha$ is related to $d_f$ via $\alpha=1+2/d_f$~\cite{makse1998pre,bunde2012fractals,makse1995nature}. This relationship was verified for obesity clusters~\cite{gallos2012collective} but does not hold well in our case. This happens because differently from the percolation model results, the probability distribution $P(A)$ is not in good agreement with a power-law function. In spite of a lack of a quantitative agreement, these models may help in understanding the mechanisms underlying the formation of the spatial patterns. In particular, these models explain that interactions among units (cities) are essential for the emergence of non-trivial spatial patterns. If these interactions are missing, the spatial structures would be formed in a randomly uniform fashion. In the case of illiteracy, this comparison suggests that the individual ties among people forming urban systems represent a key ingredient for explaining the similar illiteracy rates among neighboring cities.

\begin{table}[!ht]
\centering
\caption{Likelihood ratio tests comparing the power-law distribution against three alternatives hypotheses. The values in this table represent the test statistic (log-likelihood ratio) and values within brackets are the $p$-values of the tests. Negative log-likelihood ratio values indicate that data is likely to follow the alternative distribution, while positive values indicate that the power-law is likely preferred. The statistical significance of each choice is determined by the $p$-values.\vspace{0.3cm}}\label{table:lr}
\begin{tabular}{l|c|c|c}
\hline
\multirow{2}{*}{\bf year~}  & \multicolumn{3}{c}{\bf Alternative distributions} \\ \cline{2-4}
 & \bf lognormal & \bf stretched exponential & \bf truncated power law \\ \hline
1991 & $-0.63\;(0.53)$ & $-0.69\;(0.35)$ & $-0.99\;(0.49)$ \\ \hline
2000 & $-0.88\;(0.38)$ & $-0.94\;(0.15)$ & $-1.21\;(0.29)$ \\ \hline
2010 & $-0.69\;(0.49)$ & $-0.73\;(0.25)$ & $-1.08\;(0.40)$ \\ \hline
\end{tabular}
\end{table}

\section*{Conclusions}
We have studied the spatial patterns of the incidence of illiteracy in Brazilian cities. Our results revealed that illiteracy rates have long-range correlations and non-trivial clustering structures very similar to those observed for the spreading of diseases such as obesity~\cite{gallos2012collective} and dengue fever~\cite{antonio2017spatial}. We have also argued that these spatial patterns can be described by percolation models with long-range correlations at criticality. Following the conceptual framework  of Christakis and Fowler~\cite{christakis2009connected,christakis2007spread,christakis2008collective,fowler2008dynamic}, our results indicate that the prevalence of illiteracy in urban systems is similarly structured. Further, the methodology indicates structural similarities between what is classed as endemic, here used to describe a condition with continuous relatively stable presence, or epidemic, a condition that is rapidly increasing. As such, the methodology may be useful for studying diseases such as tuberculosis and sexually transmitted diseases in areas where they have a low prevalence but are endemic. Hypothesizing a disease-like vector for illiteracy may be controversial, however, examples exist of unexpected conditions with proven or proposed infectious components including obesity~\cite{ridaura2013gut}, schizophrenia~\cite{yolken2009toxoplasma}, and ulcers~\cite{forbes1994duodenal}. Illiteracy may be a ``purely'' socially transmitted condition but its associations with diabetes, hypertension, depression, schizophrenia, smoking, violence, and reduced life expectancy  make it an important target for improving public health outcomes. In either case, the spatial patterns are quite robust over time, supporting the hypothesis that endemic illiteracy in cities behaves like a transmissible disease. Also, the similarities with critical phenomena suggest that the incidence of illiteracy results from a collective behavior emerging from the social and economic interactions among people. Thus, like many other physical systems at criticality, these patterns are likely to depend very weakly on individual features and choices. Naturally, this does not mean people choose or want to remain illiterate, but that there are people who have not been exposed to the minimal socioeconomic conditions for becoming literate. In this endemic context, ``being sick'' (that is, remaining illiterate) must be understood as a lack of minimal education. Our results have shown that such conditions prevail over the Brazilian population in a similar fashion to traditional transmissible diseases. This result suggests that local actions against illiteracy are unlikely to have a significant impact on illiteracy rates of the entire urban system and that global campaigns would be capable of affecting collective behavior and promote a further decline in illiteracy rates.

\section*{Materials and Methods}

\subsection*{Dataset}
The dataset employed in this study consists of the illiteracy rate (percentage of illiterate people) and the geographic location (latitude and longitude) for each Brazilian city. These data were compiled by the Brazilian Institute of Geography and Statistics~\cite{IBGE} (IBGE) during the three latest demographic census that took place in 1991, 2000, and 2010. According to the IBGE methodology, a person is considered illiterate when he/she is aged 15 years or older and cannot read and write at least a single ticket in the language he/she knows. This dataset is maintained and made freely available by the IBGE~\cite{IBGE}.

\subsection*{The spatial correlation function}
The spatial correlation function is calculated via
\begin{equation}\label{eq:corr}
C(r) = \frac{\langle (i_k - \mu(r)) (i_l - \mu(r)) \rangle_{d_{kl}=r}}{{\sigma^2(r)}}\,,
\end{equation}
where $i_k$ and $i_l$ stand for the illiteracy rate in the $k$-th and $l$-th cities, $\mu(r)$ is the average of the illiteracy rate over all cities separated by $r$ kilometers, $\sigma^2(r)$ represents the variance of the same quantity, and $\langle \dots \rangle_{d_{kj}=r}$ stands for the average value over all pair of cities separated by $r$ kilometers. Because of the discrete nature of our data, $\langle \dots \rangle_{d_{kj}=r}$ is actually carried out over all pairs of cities whose distances are within the interval $(r,r+\Delta r)$. The results presented in~\autoref{fig:2} and~\autoref{sfig2} are obtained by considering thirty log-spaced distance windows, but results are robust against different choices.



\section*{Funding}
This research was supported by CNPq (Grant No. 303250/2015-1), CAPES, and FAPESP (Grant No. 2016/16987-7)

\bibliography{illiteracy_spreading}

\begin{thebibliography}{54}%
\makeatletter
\providecommand \@ifxundefined [1]{%
 \@ifx{#1\undefined}
}%
\providecommand \@ifnum [1]{%
 \ifnum #1\expandafter \@firstoftwo
 \else \expandafter \@secondoftwo
 \fi
}%
\providecommand \@ifx [1]{%
 \ifx #1\expandafter \@firstoftwo
 \else \expandafter \@secondoftwo
 \fi
}%
\providecommand \natexlab [1]{#1}%
\providecommand \enquote  [1]{``#1''}%
\providecommand \bibnamefont  [1]{#1}%
\providecommand \bibfnamefont [1]{#1}%
\providecommand \citenamefont [1]{#1}%
\providecommand \href@noop [0]{\@secondoftwo}%
\providecommand \href [0]{\begingroup \@sanitize@url \@href}%
\providecommand \@href[1]{\@@startlink{#1}\@@href}%
\providecommand \@@href[1]{\endgroup#1\@@endlink}%
\providecommand \@sanitize@url [0]{\catcode `\\12\catcode `\$12\catcode
  `\&12\catcode `\#12\catcode `\^12\catcode `\_12\catcode `\%12\relax}%
\providecommand \@@startlink[1]{}%
\providecommand \@@endlink[0]{}%
\providecommand \url  [0]{\begingroup\@sanitize@url \@url }%
\providecommand \@url [1]{\endgroup\@href {#1}{\urlprefix }}%
\providecommand \urlprefix  [0]{URL }%
\providecommand \Eprint [0]{\href }%
\providecommand \doibase [0]{http://dx.doi.org/}%
\providecommand \selectlanguage [0]{\@gobble}%
\providecommand \bibinfo  [0]{\@secondoftwo}%
\providecommand \bibfield  [0]{\@secondoftwo}%
\providecommand \translation [1]{[#1]}%
\providecommand \BibitemOpen [0]{}%
\providecommand \bibitemStop [0]{}%
\providecommand \bibitemNoStop [0]{.\EOS\space}%
\providecommand \EOS [0]{\spacefactor3000\relax}%
\providecommand \BibitemShut  [1]{\csname bibitem#1\endcsname}%
\let\auto@bib@innerbib\@empty
\bibitem [{UNE()}]{UNESCO}%
  \BibitemOpen
  \href@noop {} {\enquote {\bibinfo {title} {{United Nations Educational,
  Scientific, and Cultural Organization (UNESCO). Reading the past, writing the
  future: Fifty years of promoting literacy. 2017.}}}\ }\bibinfo {howpublished}
  {{Available:} \url{http://data.worldbank.org/indicator/SP.URB.TOTL.IN.ZS}},\
  \bibinfo {note} {{Accessed:} 13 Jan 2017}\BibitemShut {NoStop}%
\bibitem [{\citenamefont {Kutner}, \citenamefont {Greenberg},\ and\
  \citenamefont {Baer}(2006)}]{kutner2006first}%
  \BibitemOpen
  \bibfield  {author} {\bibinfo {author} {\bibfnamefont {M.}~\bibnamefont
  {Kutner}}, \bibinfo {author} {\bibfnamefont {E.}~\bibnamefont {Greenberg}}, \
  and\ \bibinfo {author} {\bibfnamefont {J.}~\bibnamefont {Baer}},\ }\href@noop
  {} {\enquote {\bibinfo {title} {A first look at the literacy of america's
  adults in the 21st century.}}\ }\bibinfo {howpublished} {{Available:}
  \url{https://nces.ed.gov/NAAL/PDF/2006470.PDF}} (\bibinfo {year} {2006}),\
  \bibinfo {note} {{Accessed:} 13 Jan 2017}\BibitemShut {NoStop}%
\bibitem [{\citenamefont {DeWalt}\ \emph {et~al.}(2004)\citenamefont {DeWalt},
  \citenamefont {Berkman}, \citenamefont {Sheridan}, \citenamefont {Lohr},\
  and\ \citenamefont {Pignone}}]{dewalt2004literacy}%
  \BibitemOpen
  \bibfield  {author} {\bibinfo {author} {\bibfnamefont {D.~A.}\ \bibnamefont
  {DeWalt}}, \bibinfo {author} {\bibfnamefont {N.~D.}\ \bibnamefont {Berkman}},
  \bibinfo {author} {\bibfnamefont {S.}~\bibnamefont {Sheridan}}, \bibinfo
  {author} {\bibfnamefont {K.~N.}\ \bibnamefont {Lohr}}, \ and\ \bibinfo
  {author} {\bibfnamefont {M.~P.}\ \bibnamefont {Pignone}},\ }\bibfield
  {title} {\enquote {\bibinfo {title} {Literacy and health outcomes: a
  systematic review of the literature},}\ }\href {\doibase
  10.1111/j.1525-1497.2004.40153.x} {\bibfield  {journal} {\bibinfo  {journal}
  {Journal of General Internal Medicine}\ }\textbf {\bibinfo {volume} {19}},\
  \bibinfo {pages} {1228--1239} (\bibinfo {year} {2004})}\BibitemShut {NoStop}%
\bibitem [{\citenamefont {Berkman}\ \emph {et~al.}(2011)\citenamefont
  {Berkman}, \citenamefont {Sheridan}, \citenamefont {Donahue}, \citenamefont
  {Halpern},\ and\ \citenamefont {Crotty}}]{berkman2011low}%
  \BibitemOpen
  \bibfield  {author} {\bibinfo {author} {\bibfnamefont {N.~D.}\ \bibnamefont
  {Berkman}}, \bibinfo {author} {\bibfnamefont {S.~L.}\ \bibnamefont
  {Sheridan}}, \bibinfo {author} {\bibfnamefont {K.~E.}\ \bibnamefont
  {Donahue}}, \bibinfo {author} {\bibfnamefont {D.~J.}\ \bibnamefont
  {Halpern}}, \ and\ \bibinfo {author} {\bibfnamefont {K.}~\bibnamefont
  {Crotty}},\ }\bibfield  {title} {\enquote {\bibinfo {title} {Low health
  literacy and health outcomes: an updated systematic review},}\ }\href
  {\doibase 10.7326/0003-4819-155-2-201107190-00005} {\bibfield  {journal}
  {\bibinfo  {journal} {Annals of Internal Medicine}\ }\textbf {\bibinfo
  {volume} {155}},\ \bibinfo {pages} {97--107} (\bibinfo {year}
  {2011})}\BibitemShut {NoStop}%
\bibitem [{\citenamefont {Schillinger}\ \emph {et~al.}(2002)\citenamefont
  {Schillinger}, \citenamefont {Grumbach}, \citenamefont {Piette},
  \citenamefont {Wang}, \citenamefont {Osmond}, \citenamefont {Daher},
  \citenamefont {Palacios}, \citenamefont {Sullivan},\ and\ \citenamefont
  {Bindman}}]{schillinger2002association}%
  \BibitemOpen
  \bibfield  {author} {\bibinfo {author} {\bibfnamefont {D.}~\bibnamefont
  {Schillinger}}, \bibinfo {author} {\bibfnamefont {K.}~\bibnamefont
  {Grumbach}}, \bibinfo {author} {\bibfnamefont {J.}~\bibnamefont {Piette}},
  \bibinfo {author} {\bibfnamefont {F.}~\bibnamefont {Wang}}, \bibinfo {author}
  {\bibfnamefont {D.}~\bibnamefont {Osmond}}, \bibinfo {author} {\bibfnamefont
  {C.}~\bibnamefont {Daher}}, \bibinfo {author} {\bibfnamefont
  {J.}~\bibnamefont {Palacios}}, \bibinfo {author} {\bibfnamefont {G.~D.}\
  \bibnamefont {Sullivan}}, \ and\ \bibinfo {author} {\bibfnamefont {A.~B.}\
  \bibnamefont {Bindman}},\ }\bibfield  {title} {\enquote {\bibinfo {title}
  {Association of health literacy with diabetes outcomes},}\ }\href {\doibase
  10.1001/jama.288.4.475} {\bibfield  {journal} {\bibinfo  {journal} {{JAMA}}\
  }\textbf {\bibinfo {volume} {288}},\ \bibinfo {pages} {475--482} (\bibinfo
  {year} {2002})}\BibitemShut {NoStop}%
\bibitem [{\citenamefont {Williams}\ \emph {et~al.}(1998)\citenamefont
  {Williams}, \citenamefont {Baker}, \citenamefont {Parker},\ and\
  \citenamefont {Nurss}}]{williams1998relationship}%
  \BibitemOpen
  \bibfield  {author} {\bibinfo {author} {\bibfnamefont {M.~V.}\ \bibnamefont
  {Williams}}, \bibinfo {author} {\bibfnamefont {D.~W.}\ \bibnamefont {Baker}},
  \bibinfo {author} {\bibfnamefont {R.~M.}\ \bibnamefont {Parker}}, \ and\
  \bibinfo {author} {\bibfnamefont {J.~R.}\ \bibnamefont {Nurss}},\ }\bibfield
  {title} {\enquote {\bibinfo {title} {Relationship of functional health
  literacy to patients' knowledge of their chronic disease: a study of patients
  with hypertension and diabetes},}\ }\href {\doibase
  10.1001/archinte.158.2.166} {\bibfield  {journal} {\bibinfo  {journal}
  {Archives of Internal Medicine}\ }\textbf {\bibinfo {volume} {158}},\
  \bibinfo {pages} {166--172} (\bibinfo {year} {1998})}\BibitemShut {NoStop}%
\bibitem [{\citenamefont {Gazmararian}\ \emph {et~al.}(2000)\citenamefont
  {Gazmararian}, \citenamefont {Baker}, \citenamefont {Parker},\ and\
  \citenamefont {Blazer}}]{gazmararian2000multivariate}%
  \BibitemOpen
  \bibfield  {author} {\bibinfo {author} {\bibfnamefont {J.}~\bibnamefont
  {Gazmararian}}, \bibinfo {author} {\bibfnamefont {D.}~\bibnamefont {Baker}},
  \bibinfo {author} {\bibfnamefont {R.}~\bibnamefont {Parker}}, \ and\ \bibinfo
  {author} {\bibfnamefont {D.~G.}\ \bibnamefont {Blazer}},\ }\bibfield  {title}
  {\enquote {\bibinfo {title} {A multivariate analysis of factors associated
  with depression: evaluating the role of health literacy as a potential
  contributor},}\ }\href {\doibase 10.1001/archinte.160.21.3307} {\bibfield
  {journal} {\bibinfo  {journal} {Archives of Internal Medicine}\ }\textbf
  {\bibinfo {volume} {160}},\ \bibinfo {pages} {3307--3314} (\bibinfo {year}
  {2000})}\BibitemShut {NoStop}%
\bibitem [{\citenamefont {Liu}\ \emph {et~al.}(2013)\citenamefont {Liu},
  \citenamefont {Song}, \citenamefont {Chen}, \citenamefont {Buka},
  \citenamefont {Zhang}, \citenamefont {Pang},\ and\ \citenamefont
  {Zheng}}]{liu2013illiteracy}%
  \BibitemOpen
  \bibfield  {author} {\bibinfo {author} {\bibfnamefont {T.}~\bibnamefont
  {Liu}}, \bibinfo {author} {\bibfnamefont {X.}~\bibnamefont {Song}}, \bibinfo
  {author} {\bibfnamefont {G.}~\bibnamefont {Chen}}, \bibinfo {author}
  {\bibfnamefont {S.~L.}\ \bibnamefont {Buka}}, \bibinfo {author}
  {\bibfnamefont {L.}~\bibnamefont {Zhang}}, \bibinfo {author} {\bibfnamefont
  {L.}~\bibnamefont {Pang}}, \ and\ \bibinfo {author} {\bibfnamefont
  {X.}~\bibnamefont {Zheng}},\ }\bibfield  {title} {\enquote {\bibinfo {title}
  {Illiteracy and schizophrenia in {China}: a population-based survey},}\
  }\href {\doibase 10.1007/s00127-012-0552-3} {\bibfield  {journal} {\bibinfo
  {journal} {Social Psychiatry And Psychiatric Epidemiology}\ }\textbf
  {\bibinfo {volume} {48}},\ \bibinfo {pages} {455--464} (\bibinfo {year}
  {2013})}\BibitemShut {NoStop}%
\bibitem [{\citenamefont {Gavarasana}, \citenamefont {Gorty},\ and\
  \citenamefont {Allam}(1992)}]{gavarasana1992illiteracy}%
  \BibitemOpen
  \bibfield  {author} {\bibinfo {author} {\bibfnamefont {S.}~\bibnamefont
  {Gavarasana}}, \bibinfo {author} {\bibfnamefont {P.~V.}\ \bibnamefont
  {Gorty}}, \ and\ \bibinfo {author} {\bibfnamefont {A.}~\bibnamefont
  {Allam}},\ }\bibfield  {title} {\enquote {\bibinfo {title} {Illiteracy,
  ignorance, and willingness to quit smoking among villagers in {India}},}\
  }\href {\doibase 10.1111/j.1349-7006.1992.tb00112.x} {\bibfield  {journal}
  {\bibinfo  {journal} {Cancer Science}\ }\textbf {\bibinfo {volume} {83}},\
  \bibinfo {pages} {340--343} (\bibinfo {year} {1992})}\BibitemShut {NoStop}%
\bibitem [{\citenamefont {Davis}\ \emph {et~al.}(1999)\citenamefont {Davis},
  \citenamefont {Byrd}, \citenamefont {Arnold}, \citenamefont {Auinger},\ and\
  \citenamefont {Bocchini}}]{davis1999low}%
  \BibitemOpen
  \bibfield  {author} {\bibinfo {author} {\bibfnamefont {T.~C.}\ \bibnamefont
  {Davis}}, \bibinfo {author} {\bibfnamefont {R.~S.}\ \bibnamefont {Byrd}},
  \bibinfo {author} {\bibfnamefont {C.~L.}\ \bibnamefont {Arnold}}, \bibinfo
  {author} {\bibfnamefont {P.}~\bibnamefont {Auinger}}, \ and\ \bibinfo
  {author} {\bibfnamefont {J.~A.}\ \bibnamefont {Bocchini}},\ }\bibfield
  {title} {\enquote {\bibinfo {title} {Low literacy and violence among
  adolescents in a summer sports program},}\ }\href {\doibase
  10.1016/S1054-139X(98)00148-7} {\bibfield  {journal} {\bibinfo  {journal}
  {Journal of Adolescent Health}\ }\textbf {\bibinfo {volume} {24}},\ \bibinfo
  {pages} {403--411} (\bibinfo {year} {1999})}\BibitemShut {NoStop}%
\bibitem [{rep(2008)}]{reportpolice}%
  \BibitemOpen
  \href@noop {} {\enquote {\bibinfo {title} {Literacy awareness resource manual
  for police. {Literacy and policing project of the {Canadian} association of
  chiefs of police}},}\ }\bibinfo {howpublished} {{Available:}
  \url{http://policeabc.ca/images/stories/CACP\_workbook\_EN\_FINAL.pdf}}
  (\bibinfo {year} {2008}),\ \bibinfo {note} {{Accessed:} 13 Jan
  2018}\BibitemShut {NoStop}%
\bibitem [{\citenamefont {Messias}(2003)}]{messias2003income}%
  \BibitemOpen
  \bibfield  {author} {\bibinfo {author} {\bibfnamefont {E.}~\bibnamefont
  {Messias}},\ }\bibfield  {title} {\enquote {\bibinfo {title} {Income
  inequality, illiteracy rate, and life expectancy in {Brazil}},}\ }\href
  {\doibase 10.2105/AJPH.93.8.1294} {\bibfield  {journal} {\bibinfo  {journal}
  {American Journal of Public Health}\ }\textbf {\bibinfo {volume} {93}},\
  \bibinfo {pages} {1294--1296} (\bibinfo {year} {2003})}\BibitemShut {NoStop}%
\bibitem [{\citenamefont {Baker}\ \emph {et~al.}(1997)\citenamefont {Baker},
  \citenamefont {Parker}, \citenamefont {Williams}, \citenamefont {Clark},\
  and\ \citenamefont {Nurss}}]{baker1997relationship}%
  \BibitemOpen
  \bibfield  {author} {\bibinfo {author} {\bibfnamefont {D.~W.}\ \bibnamefont
  {Baker}}, \bibinfo {author} {\bibfnamefont {R.~M.}\ \bibnamefont {Parker}},
  \bibinfo {author} {\bibfnamefont {M.~V.}\ \bibnamefont {Williams}}, \bibinfo
  {author} {\bibfnamefont {W.~S.}\ \bibnamefont {Clark}}, \ and\ \bibinfo
  {author} {\bibfnamefont {J.}~\bibnamefont {Nurss}},\ }\bibfield  {title}
  {\enquote {\bibinfo {title} {The relationship of patient reading ability to
  self-reported health and use of health services},}\ }\href {\doibase
  10.2105/ajph.87.6.1027} {\bibfield  {journal} {\bibinfo  {journal} {American
  Journal Of Public Health}\ }\textbf {\bibinfo {volume} {87}},\ \bibinfo
  {pages} {1027--1030} (\bibinfo {year} {1997})}\BibitemShut {NoStop}%
\bibitem [{\citenamefont {Roman}(2004)}]{roman2004illiteracy}%
  \BibitemOpen
  \bibfield  {author} {\bibinfo {author} {\bibfnamefont {S.~P.}\ \bibnamefont
  {Roman}},\ }\bibfield  {title} {\enquote {\bibinfo {title} {Illiteracy and
  older adults: Individual and societal implications},}\ }\href {\doibase
  10.1080/03601270490266257} {\bibfield  {journal} {\bibinfo  {journal}
  {Educational Gerontology}\ }\textbf {\bibinfo {volume} {30}},\ \bibinfo
  {pages} {79--93} (\bibinfo {year} {2004})}\BibitemShut {NoStop}%
\bibitem [{\citenamefont {Costa}(1988)}]{costa1988adult}%
  \BibitemOpen
  \bibfield  {author} {\bibinfo {author} {\bibfnamefont {M.}~\bibnamefont
  {Costa}},\ }\href@noop {} {\emph {\bibinfo {title} {Adult literacy/illiteracy
  in the {United States}: A handbook for reference and research}}}\ (\bibinfo
  {publisher} {ABC/Clio, Santa Barbara},\ \bibinfo {year} {1988})\BibitemShut
  {NoStop}%
\bibitem [{\citenamefont {Alves}\ \emph {et~al.}(2013)\citenamefont {Alves},
  \citenamefont {Ribeiro}, \citenamefont {Lenzi},\ and\ \citenamefont
  {Mendes}}]{alves2013distance}%
  \BibitemOpen
  \bibfield  {author} {\bibinfo {author} {\bibfnamefont {L.~G.~A.}\
  \bibnamefont {Alves}}, \bibinfo {author} {\bibfnamefont {H.~V.}\ \bibnamefont
  {Ribeiro}}, \bibinfo {author} {\bibfnamefont {E.~K.}\ \bibnamefont {Lenzi}},
  \ and\ \bibinfo {author} {\bibfnamefont {R.~S.}\ \bibnamefont {Mendes}},\
  }\bibfield  {title} {\enquote {\bibinfo {title} {{Distance to the scaling
  law: A useful approach for unveiling relationships between crime and urban
  metrics}},}\ }\href {\doibase 10.1371/journal.pone.0069580} {\bibfield
  {journal} {\bibinfo  {journal} {PLoS ONE}\ }\textbf {\bibinfo {volume} {8}},\
  \bibinfo {pages} {e0069580} (\bibinfo {year} {2013})}\BibitemShut {NoStop}%
\bibitem [{\citenamefont {Alves}\ \emph
  {et~al.}(2015{\natexlab{a}})\citenamefont {Alves}, \citenamefont {Mendes},
  \citenamefont {Lenzi},\ and\ \citenamefont {Ribeiro}}]{alves2015scale}%
  \BibitemOpen
  \bibfield  {author} {\bibinfo {author} {\bibfnamefont {L.~G.~A.}\
  \bibnamefont {Alves}}, \bibinfo {author} {\bibfnamefont {R.~S.}\ \bibnamefont
  {Mendes}}, \bibinfo {author} {\bibfnamefont {E.~K.}\ \bibnamefont {Lenzi}}, \
  and\ \bibinfo {author} {\bibfnamefont {H.~V.}\ \bibnamefont {Ribeiro}},\
  }\bibfield  {title} {\enquote {\bibinfo {title} {Scale-adjusted metrics for
  predicting the evolution of urban indicators and quantifying the performance
  of cities},}\ }\href {\doibase 10.1371/journal.pone.0134862} {\bibfield
  {journal} {\bibinfo  {journal} {PLoS ONE}\ }\textbf {\bibinfo {volume}
  {10}},\ \bibinfo {pages} {e0134862} (\bibinfo {year}
  {2015}{\natexlab{a}})}\BibitemShut {NoStop}%
\bibitem [{\citenamefont {Christakis}\ and\ \citenamefont
  {Fowler}(2009)}]{christakis2009connected}%
  \BibitemOpen
  \bibfield  {author} {\bibinfo {author} {\bibfnamefont {N.}~\bibnamefont
  {Christakis}}\ and\ \bibinfo {author} {\bibfnamefont {J.}~\bibnamefont
  {Fowler}},\ }\href@noop {} {\emph {\bibinfo {title} {Connected: The
  Surprising Power of Our Social Networks and how They Shape Our Lives}}}\
  (\bibinfo  {publisher} {Little, Brown and Company, New York},\ \bibinfo
  {year} {2009})\BibitemShut {NoStop}%
\bibitem [{\citenamefont {Christakis}\ and\ \citenamefont
  {Fowler}(2007)}]{christakis2007spread}%
  \BibitemOpen
  \bibfield  {author} {\bibinfo {author} {\bibfnamefont {N.~A.}\ \bibnamefont
  {Christakis}}\ and\ \bibinfo {author} {\bibfnamefont {J.~H.}\ \bibnamefont
  {Fowler}},\ }\bibfield  {title} {\enquote {\bibinfo {title} {The spread of
  obesity in a large social network over 32 years},}\ }\href {\doibase
  10.1056/NEJMsa066082} {\bibfield  {journal} {\bibinfo  {journal} {New England
  Journal of Medicine}\ }\textbf {\bibinfo {volume} {2007}},\ \bibinfo {pages}
  {370--379} (\bibinfo {year} {2007})}\BibitemShut {NoStop}%
\bibitem [{\citenamefont {Christakis}\ and\ \citenamefont
  {Fowler}(2008)}]{christakis2008collective}%
  \BibitemOpen
  \bibfield  {author} {\bibinfo {author} {\bibfnamefont {N.~A.}\ \bibnamefont
  {Christakis}}\ and\ \bibinfo {author} {\bibfnamefont {J.~H.}\ \bibnamefont
  {Fowler}},\ }\bibfield  {title} {\enquote {\bibinfo {title} {The collective
  dynamics of smoking in a large social network},}\ }\href {\doibase
  10.1056/NEJMsa0706154} {\bibfield  {journal} {\bibinfo  {journal} {New
  England Journal of Medicine}\ }\textbf {\bibinfo {volume} {358}},\ \bibinfo
  {pages} {2249--2258} (\bibinfo {year} {2008})}\BibitemShut {NoStop}%
\bibitem [{\citenamefont {Fowler}\ and\ \citenamefont
  {Christakis}(2008)}]{fowler2008dynamic}%
  \BibitemOpen
  \bibfield  {author} {\bibinfo {author} {\bibfnamefont {J.~H.}\ \bibnamefont
  {Fowler}}\ and\ \bibinfo {author} {\bibfnamefont {N.~A.}\ \bibnamefont
  {Christakis}},\ }\bibfield  {title} {\enquote {\bibinfo {title} {Dynamic
  spread of happiness in a large social network: longitudinal analysis over 20
  years in the {Framingham Heart Study}},}\ }\href {\doibase 10.1136/bmj.a2338}
  {\bibfield  {journal} {\bibinfo  {journal} {{BMJ}}\ }\textbf {\bibinfo
  {volume} {337}},\ \bibinfo {pages} {a2338} (\bibinfo {year}
  {2008})}\BibitemShut {NoStop}%
\bibitem [{\citenamefont {Kulldorff}\ and\ \citenamefont
  {Nagarwalla}(1995)}]{kulldorff1995spatial}%
  \BibitemOpen
  \bibfield  {author} {\bibinfo {author} {\bibfnamefont {M.}~\bibnamefont
  {Kulldorff}}\ and\ \bibinfo {author} {\bibfnamefont {N.}~\bibnamefont
  {Nagarwalla}},\ }\bibfield  {title} {\enquote {\bibinfo {title} {Spatial
  disease clusters: detection and inference},}\ }\href {\doibase
  10.1002/sim.4780140809} {\bibfield  {journal} {\bibinfo  {journal}
  {Statistics in Medicine}\ }\textbf {\bibinfo {volume} {14}},\ \bibinfo
  {pages} {799--810} (\bibinfo {year} {1995})}\BibitemShut {NoStop}%
\bibitem [{\citenamefont {Tilman}\ and\ \citenamefont
  {Kareiva}(1997)}]{tilman1997spatial}%
  \BibitemOpen
  \bibfield  {author} {\bibinfo {author} {\bibfnamefont {D.}~\bibnamefont
  {Tilman}}\ and\ \bibinfo {author} {\bibfnamefont {P.~M.}\ \bibnamefont
  {Kareiva}},\ }\href@noop {} {\emph {\bibinfo {title} {Spatial ecology: the
  role of space in population dynamics and interspecific interactions}}},\
  Vol.~\bibinfo {volume} {30}\ (\bibinfo  {publisher} {Princeton University
  Press, Chichester},\ \bibinfo {year} {1997})\BibitemShut {NoStop}%
\bibitem [{\citenamefont {Grenfell}, \citenamefont {Bj{ø}rnstad},\ and\
  \citenamefont {Finkenstädt}(2002)}]{grenfell2002dynamics}%
  \BibitemOpen
  \bibfield  {author} {\bibinfo {author} {\bibfnamefont {B.~T.}\ \bibnamefont
  {Grenfell}}, \bibinfo {author} {\bibfnamefont {O.~N.}\ \bibnamefont
  {Bj{ø}rnstad}}, \ and\ \bibinfo {author} {\bibfnamefont {B.~F.}\
  \bibnamefont {Finkenstädt}},\ }\bibfield  {title} {\enquote {\bibinfo
  {title} {Dynamics of measles epidemics: scaling noise, determinism, and
  predictability with the {TSIR} model},}\ }\href {\doibase
  10.1890/0012-9615(2002)072[0185:domesn]2.0.co;2} {\bibfield  {journal}
  {\bibinfo  {journal} {Ecological Monographs}\ }\textbf {\bibinfo {volume}
  {72}},\ \bibinfo {pages} {185--202} (\bibinfo {year} {2002})}\BibitemShut
  {NoStop}%
\bibitem [{\citenamefont {Viboud}\ \emph {et~al.}(2006)\citenamefont {Viboud},
  \citenamefont {Bj{ø}rnstad}, \citenamefont {Smith}, \citenamefont
  {Simonsen}, \citenamefont {Miller},\ and\ \citenamefont
  {Grenfell}}]{viboud2006synchrony}%
  \BibitemOpen
  \bibfield  {author} {\bibinfo {author} {\bibfnamefont {C.}~\bibnamefont
  {Viboud}}, \bibinfo {author} {\bibfnamefont {O.~N.}\ \bibnamefont
  {Bj{ø}rnstad}}, \bibinfo {author} {\bibfnamefont {D.~L.}\ \bibnamefont
  {Smith}}, \bibinfo {author} {\bibfnamefont {L.}~\bibnamefont {Simonsen}},
  \bibinfo {author} {\bibfnamefont {M.~A.}\ \bibnamefont {Miller}}, \ and\
  \bibinfo {author} {\bibfnamefont {B.~T.}\ \bibnamefont {Grenfell}},\
  }\bibfield  {title} {\enquote {\bibinfo {title} {Synchrony, waves, and
  spatial hierarchies in the spread of influenza},}\ }\href {\doibase
  10.1126/science.1125237} {\bibfield  {journal} {\bibinfo  {journal}
  {Science}\ }\textbf {\bibinfo {volume} {312}},\ \bibinfo {pages} {447--451}
  (\bibinfo {year} {2006})}\BibitemShut {NoStop}%
\bibitem [{\citenamefont {Riley}(2007)}]{riley2007large}%
  \BibitemOpen
  \bibfield  {author} {\bibinfo {author} {\bibfnamefont {S.}~\bibnamefont
  {Riley}},\ }\bibfield  {title} {\enquote {\bibinfo {title} {Large-scale
  spatial-transmission models of infectious disease},}\ }\href {\doibase
  10.1126/science.1134695} {\bibfield  {journal} {\bibinfo  {journal}
  {Science}\ }\textbf {\bibinfo {volume} {316}},\ \bibinfo {pages} {1298--1301}
  (\bibinfo {year} {2007})}\BibitemShut {NoStop}%
\bibitem [{\citenamefont {Chowell}\ \emph {et~al.}(2008)\citenamefont
  {Chowell}, \citenamefont {Bettencourt}, \citenamefont {Johnson},
  \citenamefont {Alonso},\ and\ \citenamefont {Viboud}}]{chowell20081918}%
  \BibitemOpen
  \bibfield  {author} {\bibinfo {author} {\bibfnamefont {G.}~\bibnamefont
  {Chowell}}, \bibinfo {author} {\bibfnamefont {L.~M.}\ \bibnamefont
  {Bettencourt}}, \bibinfo {author} {\bibfnamefont {N.}~\bibnamefont
  {Johnson}}, \bibinfo {author} {\bibfnamefont {W.~J.}\ \bibnamefont {Alonso}},
  \ and\ \bibinfo {author} {\bibfnamefont {C.}~\bibnamefont {Viboud}},\
  }\bibfield  {title} {\enquote {\bibinfo {title} {The 1918--1919 influenza
  pandemic in {England} and {Wales}: spatial patterns in transmissibility and
  mortality impact},}\ }\href {\doibase 10.1098/rspb.2007.1477} {\bibfield
  {journal} {\bibinfo  {journal} {Proceedings of the Royal Society of London B:
  Biological Sciences}\ }\textbf {\bibinfo {volume} {275}},\ \bibinfo {pages}
  {501--509} (\bibinfo {year} {2008})}\BibitemShut {NoStop}%
\bibitem [{\citenamefont {Pitzer}\ \emph {et~al.}(2009)\citenamefont {Pitzer},
  \citenamefont {Viboud}, \citenamefont {Simonsen}, \citenamefont {Steiner},
  \citenamefont {Panozzo}, \citenamefont {Alonso}, \citenamefont {Miller},
  \citenamefont {Glass}, \citenamefont {Glasser}, \citenamefont {Parashar}
  \emph {et~al.}}]{pitzer2009demographic}%
  \BibitemOpen
  \bibfield  {author} {\bibinfo {author} {\bibfnamefont {V.~E.}\ \bibnamefont
  {Pitzer}}, \bibinfo {author} {\bibfnamefont {C.}~\bibnamefont {Viboud}},
  \bibinfo {author} {\bibfnamefont {L.}~\bibnamefont {Simonsen}}, \bibinfo
  {author} {\bibfnamefont {C.}~\bibnamefont {Steiner}}, \bibinfo {author}
  {\bibfnamefont {C.~A.}\ \bibnamefont {Panozzo}}, \bibinfo {author}
  {\bibfnamefont {W.~J.}\ \bibnamefont {Alonso}}, \bibinfo {author}
  {\bibfnamefont {M.~A.}\ \bibnamefont {Miller}}, \bibinfo {author}
  {\bibfnamefont {R.~I.}\ \bibnamefont {Glass}}, \bibinfo {author}
  {\bibfnamefont {J.~W.}\ \bibnamefont {Glasser}}, \bibinfo {author}
  {\bibfnamefont {U.~D.}\ \bibnamefont {Parashar}},  \emph {et~al.},\
  }\bibfield  {title} {\enquote {\bibinfo {title} {Demographic variability,
  vaccination, and the spatiotemporal dynamics of rotavirus epidemics},}\
  }\href {\doibase 10.1126/science.1172330} {\bibfield  {journal} {\bibinfo
  {journal} {Science}\ }\textbf {\bibinfo {volume} {325}},\ \bibinfo {pages}
  {290--294} (\bibinfo {year} {2009})}\BibitemShut {NoStop}%
\bibitem [{\citenamefont {Balcan}\ \emph {et~al.}(2009)\citenamefont {Balcan},
  \citenamefont {Colizza}, \citenamefont {Gonçalves}, \citenamefont {Hu},
  \citenamefont {Ramasco},\ and\ \citenamefont
  {Vespignani}}]{balcan2009multiscale}%
  \BibitemOpen
  \bibfield  {author} {\bibinfo {author} {\bibfnamefont {D.}~\bibnamefont
  {Balcan}}, \bibinfo {author} {\bibfnamefont {V.}~\bibnamefont {Colizza}},
  \bibinfo {author} {\bibfnamefont {B.}~\bibnamefont {Gonçalves}}, \bibinfo
  {author} {\bibfnamefont {H.}~\bibnamefont {Hu}}, \bibinfo {author}
  {\bibfnamefont {J.~J.}\ \bibnamefont {Ramasco}}, \ and\ \bibinfo {author}
  {\bibfnamefont {A.}~\bibnamefont {Vespignani}},\ }\bibfield  {title}
  {\enquote {\bibinfo {title} {Multiscale mobility networks and the spatial
  spreading of infectious diseases},}\ }\href {\doibase
  10.1073/pnas.0906910106} {\bibfield  {journal} {\bibinfo  {journal}
  {Proceedings of the National Academy of Sciences}\ }\textbf {\bibinfo
  {volume} {106}},\ \bibinfo {pages} {21484--21489} (\bibinfo {year}
  {2009})}\BibitemShut {NoStop}%
\bibitem [{\citenamefont {Gallos}\ \emph {et~al.}(2012)\citenamefont {Gallos},
  \citenamefont {Barttfeld}, \citenamefont {Havlin}, \citenamefont {Sigman},\
  and\ \citenamefont {Makse}}]{gallos2012collective}%
  \BibitemOpen
  \bibfield  {author} {\bibinfo {author} {\bibfnamefont {L.~K.}\ \bibnamefont
  {Gallos}}, \bibinfo {author} {\bibfnamefont {P.}~\bibnamefont {Barttfeld}},
  \bibinfo {author} {\bibfnamefont {S.}~\bibnamefont {Havlin}}, \bibinfo
  {author} {\bibfnamefont {M.}~\bibnamefont {Sigman}}, \ and\ \bibinfo {author}
  {\bibfnamefont {H.~A.}\ \bibnamefont {Makse}},\ }\bibfield  {title} {\enquote
  {\bibinfo {title} {Collective behavior in the spatial spreading of
  obesity},}\ }\href {\doibase 10.1038/srep00454} {\bibfield  {journal}
  {\bibinfo  {journal} {Scientific Reports}\ }\textbf {\bibinfo {volume} {2}}
  (\bibinfo {year} {2012}),\ 10.1038/srep00454}\BibitemShut {NoStop}%
\bibitem [{\citenamefont {Viboud}\ \emph {et~al.}(2013)\citenamefont {Viboud},
  \citenamefont {Nelson}, \citenamefont {Tan},\ and\ \citenamefont
  {Holmes}}]{viboud2013contrasting}%
  \BibitemOpen
  \bibfield  {author} {\bibinfo {author} {\bibfnamefont {C.}~\bibnamefont
  {Viboud}}, \bibinfo {author} {\bibfnamefont {M.~I.}\ \bibnamefont {Nelson}},
  \bibinfo {author} {\bibfnamefont {Y.}~\bibnamefont {Tan}}, \ and\ \bibinfo
  {author} {\bibfnamefont {E.~C.}\ \bibnamefont {Holmes}},\ }\bibfield  {title}
  {\enquote {\bibinfo {title} {Contrasting the epidemiological and evolutionary
  dynamics of influenza spatial transmission},}\ }\href {\doibase
  10.1098/rstb.2012.0199} {\bibfield  {journal} {\bibinfo  {journal}
  {Philosophical Transactions of the Royal Society B: Biological Sciences}\
  }\textbf {\bibinfo {volume} {368}},\ \bibinfo {pages} {20120199} (\bibinfo
  {year} {2013})}\BibitemShut {NoStop}%
\bibitem [{\citenamefont {Brockmann}\ and\ \citenamefont
  {Helbing}(2013)}]{brockmann2013hidden}%
  \BibitemOpen
  \bibfield  {author} {\bibinfo {author} {\bibfnamefont {D.}~\bibnamefont
  {Brockmann}}\ and\ \bibinfo {author} {\bibfnamefont {D.}~\bibnamefont
  {Helbing}},\ }\bibfield  {title} {\enquote {\bibinfo {title} {The hidden
  geometry of complex, network-driven contagion phenomena},}\ }\href {\doibase
  10.1126/science.1245200} {\bibfield  {journal} {\bibinfo  {journal}
  {Science}\ }\textbf {\bibinfo {volume} {342}},\ \bibinfo {pages} {1337--1342}
  (\bibinfo {year} {2013})}\BibitemShut {NoStop}%
\bibitem [{\citenamefont {Gog}\ \emph {et~al.}(2014)\citenamefont {Gog},
  \citenamefont {Ballesteros}, \citenamefont {Viboud}, \citenamefont
  {Simonsen}, \citenamefont {Bjornstad}, \citenamefont {Shaman}, \citenamefont
  {Chao}, \citenamefont {Khan},\ and\ \citenamefont
  {Grenfell}}]{gog2014spatial}%
  \BibitemOpen
  \bibfield  {author} {\bibinfo {author} {\bibfnamefont {J.~R.}\ \bibnamefont
  {Gog}}, \bibinfo {author} {\bibfnamefont {S.}~\bibnamefont {Ballesteros}},
  \bibinfo {author} {\bibfnamefont {C.}~\bibnamefont {Viboud}}, \bibinfo
  {author} {\bibfnamefont {L.}~\bibnamefont {Simonsen}}, \bibinfo {author}
  {\bibfnamefont {O.~N.}\ \bibnamefont {Bjornstad}}, \bibinfo {author}
  {\bibfnamefont {J.}~\bibnamefont {Shaman}}, \bibinfo {author} {\bibfnamefont
  {D.~L.}\ \bibnamefont {Chao}}, \bibinfo {author} {\bibfnamefont
  {F.}~\bibnamefont {Khan}}, \ and\ \bibinfo {author} {\bibfnamefont {B.~T.}\
  \bibnamefont {Grenfell}},\ }\bibfield  {title} {\enquote {\bibinfo {title}
  {Spatial transmission of 2009 pandemic influenza in the {US}},}\ }\href
  {\doibase 10.1371/journal.pcbi.1003635} {\bibfield  {journal} {\bibinfo
  {journal} {PLoS Computational Biology}\ }\textbf {\bibinfo {volume} {10}},\
  \bibinfo {pages} {e1003635} (\bibinfo {year} {2014})}\BibitemShut {NoStop}%
\bibitem [{\citenamefont {Sun}\ \emph {et~al.}(2016)\citenamefont {Sun},
  \citenamefont {Jusup}, \citenamefont {Jin}, \citenamefont {Wang},\ and\
  \citenamefont {Wang}}]{sun2016pattern}%
  \BibitemOpen
  \bibfield  {author} {\bibinfo {author} {\bibfnamefont {G.-Q.}\ \bibnamefont
  {Sun}}, \bibinfo {author} {\bibfnamefont {M.}~\bibnamefont {Jusup}}, \bibinfo
  {author} {\bibfnamefont {Z.}~\bibnamefont {Jin}}, \bibinfo {author}
  {\bibfnamefont {Y.}~\bibnamefont {Wang}}, \ and\ \bibinfo {author}
  {\bibfnamefont {Z.}~\bibnamefont {Wang}},\ }\bibfield  {title} {\enquote
  {\bibinfo {title} {Pattern transitions in spatial epidemics: Mechanisms and
  emergent properties},}\ }\href {\doibase 10.1016/j.plrev.2016.08.002}
  {\bibfield  {journal} {\bibinfo  {journal} {Physics of Life Reviews}\
  }\textbf {\bibinfo {volume} {19}},\ \bibinfo {pages} {43--73} (\bibinfo
  {year} {2016})}\BibitemShut {NoStop}%
\bibitem [{\citenamefont {Antonio}\ \emph {et~al.}(2017)\citenamefont
  {Antonio}, \citenamefont {Itami}, \citenamefont {de~Picoli}, \citenamefont
  {Teixeira},\ and\ \citenamefont {dos Santos~Mendes}}]{antonio2017spatial}%
  \BibitemOpen
  \bibfield  {author} {\bibinfo {author} {\bibfnamefont {F.~J.}\ \bibnamefont
  {Antonio}}, \bibinfo {author} {\bibfnamefont {A.~S.}\ \bibnamefont {Itami}},
  \bibinfo {author} {\bibfnamefont {S.}~\bibnamefont {de~Picoli}}, \bibinfo
  {author} {\bibfnamefont {J.~J.~V.}\ \bibnamefont {Teixeira}}, \ and\ \bibinfo
  {author} {\bibfnamefont {R.}~\bibnamefont {dos Santos~Mendes}},\ }\bibfield
  {title} {\enquote {\bibinfo {title} {Spatial patterns of dengue cases in
  {Brazil}},}\ }\href {\doibase 10.1371/journal.pone.0180715} {\bibfield
  {journal} {\bibinfo  {journal} {PLoS ONE}\ }\textbf {\bibinfo {volume}
  {12}},\ \bibinfo {pages} {e0180715} (\bibinfo {year} {2017})}\BibitemShut
  {NoStop}%
\bibitem [{MEC()}]{MEC}%
  \BibitemOpen
  \href@noop {} {\enquote {\bibinfo {title} {{Ministério da Educação,
  Instituto Nacional de Estudos e Pesquisas Educacionais. Mapa do Analfabetismo
  no Brasil. 2000.}}}\ }\bibinfo {howpublished} {{Available:}
  \url{http://portal.inep.gov.br/documents/186968/485745/Mapa+do+analfabetismo+no+Brasil/a53ac9ee-c0c0-4727-b216-035c65c45e1b?version=1.3}},\
  \bibinfo {note} {{Accessed:} 13 Jan 2018}\BibitemShut {NoStop}%
\bibitem [{\citenamefont {Alves}\ \emph
  {et~al.}(2015{\natexlab{b}})\citenamefont {Alves}, \citenamefont {Lenzi},
  \citenamefont {Mendes},\ and\ \citenamefont {Ribeiro}}]{alves2015spatial}%
  \BibitemOpen
  \bibfield  {author} {\bibinfo {author} {\bibfnamefont {L.~G.~A.}\
  \bibnamefont {Alves}}, \bibinfo {author} {\bibfnamefont {E.~K.}\ \bibnamefont
  {Lenzi}}, \bibinfo {author} {\bibfnamefont {R.~S.}\ \bibnamefont {Mendes}}, \
  and\ \bibinfo {author} {\bibfnamefont {H.~V.}\ \bibnamefont {Ribeiro}},\
  }\bibfield  {title} {\enquote {\bibinfo {title} {Spatial correlations,
  clustering and percolation-like transitions in homicide crimes},}\ }\href
  {\doibase 10.1209/0295-5075/111/18002} {\bibfield  {journal} {\bibinfo
  {journal} {EPL}\ }\textbf {\bibinfo {volume} {111}},\ \bibinfo {pages}
  {18002} (\bibinfo {year} {2015}{\natexlab{b}})}\BibitemShut {NoStop}%
\bibitem [{\citenamefont {Stanley}(1971)}]{stanley1971introduction}%
  \BibitemOpen
  \bibfield  {author} {\bibinfo {author} {\bibfnamefont {H.~E.}\ \bibnamefont
  {Stanley}},\ }\href@noop {} {\emph {\bibinfo {title} {Introduction to phase
  transitions and critical phenomena}}}\ (\bibinfo  {publisher} {Oxford
  University Press, Oxford},\ \bibinfo {year} {1971})\BibitemShut {NoStop}%
\bibitem [{\citenamefont {Schneidman}\ \emph {et~al.}(2006)\citenamefont
  {Schneidman}, \citenamefont {Berry~II}, \citenamefont {Segev},\ and\
  \citenamefont {Bialek}}]{schneidman2006weak}%
  \BibitemOpen
  \bibfield  {author} {\bibinfo {author} {\bibfnamefont {E.}~\bibnamefont
  {Schneidman}}, \bibinfo {author} {\bibfnamefont {M.~J.}\ \bibnamefont
  {Berry~II}}, \bibinfo {author} {\bibfnamefont {R.}~\bibnamefont {Segev}}, \
  and\ \bibinfo {author} {\bibfnamefont {W.}~\bibnamefont {Bialek}},\
  }\bibfield  {title} {\enquote {\bibinfo {title} {Weak pairwise correlations
  imply strongly correlated network states in a neural population},}\ }\href
  {\doibase 10.1038/nature04701} {\bibfield  {journal} {\bibinfo  {journal}
  {Nature}\ }\textbf {\bibinfo {volume} {440}},\ \bibinfo {pages} {1007}
  (\bibinfo {year} {2006})}\BibitemShut {NoStop}%
\bibitem [{\citenamefont {Mora}\ and\ \citenamefont
  {Bialek}(2011)}]{mora2011biological}%
  \BibitemOpen
  \bibfield  {author} {\bibinfo {author} {\bibfnamefont {T.}~\bibnamefont
  {Mora}}\ and\ \bibinfo {author} {\bibfnamefont {W.}~\bibnamefont {Bialek}},\
  }\bibfield  {title} {\enquote {\bibinfo {title} {Are biological systems
  poised at criticality?}}\ }\href {\doibase 10.1007/s10955-011-0229-4}
  {\bibfield  {journal} {\bibinfo  {journal} {Journal of Statistical Physics}\
  }\textbf {\bibinfo {volume} {144}},\ \bibinfo {pages} {268--302} (\bibinfo
  {year} {2011})}\BibitemShut {NoStop}%
\bibitem [{\citenamefont {Cavagna}\ \emph {et~al.}(2010)\citenamefont
  {Cavagna}, \citenamefont {Cimarelli}, \citenamefont {Giardina}, \citenamefont
  {Parisi}, \citenamefont {Santagati}, \citenamefont {Stefanini},\ and\
  \citenamefont {Viale}}]{cavagna2010scale}%
  \BibitemOpen
  \bibfield  {author} {\bibinfo {author} {\bibfnamefont {A.}~\bibnamefont
  {Cavagna}}, \bibinfo {author} {\bibfnamefont {A.}~\bibnamefont {Cimarelli}},
  \bibinfo {author} {\bibfnamefont {I.}~\bibnamefont {Giardina}}, \bibinfo
  {author} {\bibfnamefont {G.}~\bibnamefont {Parisi}}, \bibinfo {author}
  {\bibfnamefont {R.}~\bibnamefont {Santagati}}, \bibinfo {author}
  {\bibfnamefont {F.}~\bibnamefont {Stefanini}}, \ and\ \bibinfo {author}
  {\bibfnamefont {M.}~\bibnamefont {Viale}},\ }\bibfield  {title} {\enquote
  {\bibinfo {title} {Scale-free correlations in starling flocks},}\ }\href
  {\doibase 10.1073/pnas.1005766107} {\bibfield  {journal} {\bibinfo  {journal}
  {Proceedings of the National Academy of Sciences}\ }\textbf {\bibinfo
  {volume} {107}},\ \bibinfo {pages} {11865--11870} (\bibinfo {year}
  {2010})}\BibitemShut {NoStop}%
\bibitem [{\citenamefont {Ester}\ \emph {et~al.}(1996)\citenamefont {Ester},
  \citenamefont {Kriegel}, \citenamefont {Sander}, \citenamefont {Xu} \emph
  {et~al.}}]{ester1996density}%
  \BibitemOpen
  \bibfield  {author} {\bibinfo {author} {\bibfnamefont {M.}~\bibnamefont
  {Ester}}, \bibinfo {author} {\bibfnamefont {H.-P.}\ \bibnamefont {Kriegel}},
  \bibinfo {author} {\bibfnamefont {J.}~\bibnamefont {Sander}}, \bibinfo
  {author} {\bibfnamefont {X.}~\bibnamefont {Xu}},  \emph {et~al.},\ }\bibfield
   {title} {\enquote {\bibinfo {title} {A density-based algorithm for
  discovering clusters in large spatial databases with noise},}\ }in\
  \href@noop {} {\emph {\bibinfo {booktitle} {Proceedings of the 2nd
  International Conference on Knowledge Discovery and Data Mining}}},\
  Vol.~\bibinfo {volume} {96}\ (\bibinfo {year} {1996})\ pp.\ \bibinfo {pages}
  {226--231}\BibitemShut {NoStop}%
\bibitem [{\citenamefont {Rozenfeld}\ \emph {et~al.}(2008)\citenamefont
  {Rozenfeld}, \citenamefont {Rybski}, \citenamefont {Andrade}, \citenamefont
  {Batty}, \citenamefont {Stanley},\ and\ \citenamefont
  {Makse}}]{rozenfeld2008laws}%
  \BibitemOpen
  \bibfield  {author} {\bibinfo {author} {\bibfnamefont {H.~D.}\ \bibnamefont
  {Rozenfeld}}, \bibinfo {author} {\bibfnamefont {D.}~\bibnamefont {Rybski}},
  \bibinfo {author} {\bibfnamefont {J.~S.}\ \bibnamefont {Andrade}}, \bibinfo
  {author} {\bibfnamefont {M.}~\bibnamefont {Batty}}, \bibinfo {author}
  {\bibfnamefont {H.~E.}\ \bibnamefont {Stanley}}, \ and\ \bibinfo {author}
  {\bibfnamefont {H.~A.}\ \bibnamefont {Makse}},\ }\bibfield  {title} {\enquote
  {\bibinfo {title} {Laws of population growth},}\ }\href {\doibase
  10.1073/pnas.0807435105} {\bibfield  {journal} {\bibinfo  {journal}
  {Proceedings of the National Academy of Sciences}\ }\textbf {\bibinfo
  {volume} {105}},\ \bibinfo {pages} {18702--18707} (\bibinfo {year}
  {2008})}\BibitemShut {NoStop}%
\bibitem [{\citenamefont {Bunde}\ and\ \citenamefont
  {Havlin}(2012)}]{bunde2012fractals}%
  \BibitemOpen
  \bibfield  {author} {\bibinfo {author} {\bibfnamefont {A.}~\bibnamefont
  {Bunde}}\ and\ \bibinfo {author} {\bibfnamefont {S.}~\bibnamefont {Havlin}},\
  }\href@noop {} {\emph {\bibinfo {title} {Fractals and Disordered Systems}}}\
  (\bibinfo  {publisher} {Springer-Verlag, Heidelberg},\ \bibinfo {year}
  {2012})\BibitemShut {NoStop}%
\bibitem [{\citenamefont {Schubert}\ \emph {et~al.}(2017)\citenamefont
  {Schubert}, \citenamefont {Sander}, \citenamefont {Ester}, \citenamefont
  {Kriegel},\ and\ \citenamefont {Xu}}]{schubert2017dbscan}%
  \BibitemOpen
  \bibfield  {author} {\bibinfo {author} {\bibfnamefont {E.}~\bibnamefont
  {Schubert}}, \bibinfo {author} {\bibfnamefont {J.}~\bibnamefont {Sander}},
  \bibinfo {author} {\bibfnamefont {M.}~\bibnamefont {Ester}}, \bibinfo
  {author} {\bibfnamefont {H.~P.}\ \bibnamefont {Kriegel}}, \ and\ \bibinfo
  {author} {\bibfnamefont {X.}~\bibnamefont {Xu}},\ }\bibfield  {title}
  {\enquote {\bibinfo {title} {Dbscan revisited, revisited: Why and how you
  should (still) use dbscan},}\ }\href {\doibase 10.1145/3068335} {\bibfield
  {journal} {\bibinfo  {journal} {ACM Transactions on Database Systems (TODS)}\
  }\textbf {\bibinfo {volume} {42}},\ \bibinfo {pages} {19} (\bibinfo {year}
  {2017})}\BibitemShut {NoStop}%
\bibitem [{Note1()}]{Note1}%
  \BibitemOpen
  \bibinfo {note} {We have employed $\varepsilon =48$~km for all fractal
  analysis, but results are very robust for $25<\varepsilon
  <75$~km.}\BibitemShut {Stop}%
\bibitem [{\citenamefont {Moreira}\ and\ \citenamefont
  {Santos}(2007)}]{moreira2007concave}%
  \BibitemOpen
  \bibfield  {author} {\bibinfo {author} {\bibfnamefont {A.~J.~C.}\
  \bibnamefont {Moreira}}\ and\ \bibinfo {author} {\bibfnamefont {M.~Y.}\
  \bibnamefont {Santos}},\ }\bibfield  {title} {\enquote {\bibinfo {title}
  {Concave hull: {A} k-nearest neighbours approach for the computation of the
  region occupied by a set of points},}\ }in\ \href@noop {} {\emph {\bibinfo
  {booktitle} {{GRAPP} 2007, Proceedings of the Second International Conference
  on Computer Graphics Theory and Applications, Barcelona, Spain, March 8-11,
  2007, Volume {GM/R}}}}\ (\bibinfo {year} {2007})\ pp.\ \bibinfo {pages}
  {61--68}\BibitemShut {NoStop}%
\bibitem [{\citenamefont {Clauset}, \citenamefont {Shalizi},\ and\
  \citenamefont {Newman}(2009)}]{clauset2009power}%
  \BibitemOpen
  \bibfield  {author} {\bibinfo {author} {\bibfnamefont {A.}~\bibnamefont
  {Clauset}}, \bibinfo {author} {\bibfnamefont {C.~R.}\ \bibnamefont
  {Shalizi}}, \ and\ \bibinfo {author} {\bibfnamefont {M.~E.}\ \bibnamefont
  {Newman}},\ }\bibfield  {title} {\enquote {\bibinfo {title} {Power-law
  distributions in empirical data},}\ }\href {\doibase 10.1137/070710111}
  {\bibfield  {journal} {\bibinfo  {journal} {SIAM review}\ }\textbf {\bibinfo
  {volume} {51}},\ \bibinfo {pages} {661--703} (\bibinfo {year}
  {2009})}\BibitemShut {NoStop}%
\bibitem [{\citenamefont {Makse}\ \emph {et~al.}(1998)\citenamefont {Makse},
  \citenamefont {Andrade}, \citenamefont {Batty}, \citenamefont {Havlin},
  \citenamefont {Stanley} \emph {et~al.}}]{makse1998pre}%
  \BibitemOpen
  \bibfield  {author} {\bibinfo {author} {\bibfnamefont {H.~A.}\ \bibnamefont
  {Makse}}, \bibinfo {author} {\bibfnamefont {J.~S.}\ \bibnamefont {Andrade}},
  \bibinfo {author} {\bibfnamefont {M.}~\bibnamefont {Batty}}, \bibinfo
  {author} {\bibfnamefont {S.}~\bibnamefont {Havlin}}, \bibinfo {author}
  {\bibfnamefont {H.~E.}\ \bibnamefont {Stanley}},  \emph {et~al.},\ }\bibfield
   {title} {\enquote {\bibinfo {title} {Modeling urban growth patterns with
  correlated percolation},}\ }\href {\doibase 10.1103/PhysRevE.58.7054}
  {\bibfield  {journal} {\bibinfo  {journal} {Physical Review E}\ }\textbf
  {\bibinfo {volume} {58}},\ \bibinfo {pages} {7054} (\bibinfo {year}
  {1998})}\BibitemShut {NoStop}%
\bibitem [{\citenamefont {Makse}, \citenamefont {Havlin},\ and\ \citenamefont
  {Stanley}(1995)}]{makse1995nature}%
  \BibitemOpen
  \bibfield  {author} {\bibinfo {author} {\bibfnamefont {H.~A.}\ \bibnamefont
  {Makse}}, \bibinfo {author} {\bibfnamefont {S.}~\bibnamefont {Havlin}}, \
  and\ \bibinfo {author} {\bibfnamefont {H.~E.}\ \bibnamefont {Stanley}},\
  }\bibfield  {title} {\enquote {\bibinfo {title} {Modelling urban growth
  patterns},}\ }\href {\doibase 10.1038/377608a0} {\bibfield  {journal}
  {\bibinfo  {journal} {Nature}\ }\textbf {\bibinfo {volume} {377}},\ \bibinfo
  {pages} {608--612} (\bibinfo {year} {1995})}\BibitemShut {NoStop}%
\bibitem [{\citenamefont {Ridaura}\ \emph {et~al.}(2013)\citenamefont
  {Ridaura}, \citenamefont {Faith}, \citenamefont {Rey}, \citenamefont {Cheng},
  \citenamefont {Duncan}, \citenamefont {Kau}, \citenamefont {Griffin},
  \citenamefont {Lombard}, \citenamefont {Henrissat}, \citenamefont {Bain}
  \emph {et~al.}}]{ridaura2013gut}%
  \BibitemOpen
  \bibfield  {author} {\bibinfo {author} {\bibfnamefont {V.~K.}\ \bibnamefont
  {Ridaura}}, \bibinfo {author} {\bibfnamefont {J.~J.}\ \bibnamefont {Faith}},
  \bibinfo {author} {\bibfnamefont {F.~E.}\ \bibnamefont {Rey}}, \bibinfo
  {author} {\bibfnamefont {J.}~\bibnamefont {Cheng}}, \bibinfo {author}
  {\bibfnamefont {A.~E.}\ \bibnamefont {Duncan}}, \bibinfo {author}
  {\bibfnamefont {A.~L.}\ \bibnamefont {Kau}}, \bibinfo {author} {\bibfnamefont
  {N.~W.}\ \bibnamefont {Griffin}}, \bibinfo {author} {\bibfnamefont
  {V.}~\bibnamefont {Lombard}}, \bibinfo {author} {\bibfnamefont
  {B.}~\bibnamefont {Henrissat}}, \bibinfo {author} {\bibfnamefont {J.~R.}\
  \bibnamefont {Bain}},  \emph {et~al.},\ }\bibfield  {title} {\enquote
  {\bibinfo {title} {Gut microbiota from twins discordant for obesity modulate
  metabolism in mice},}\ }\href {\doibase 10.1126/science.1241214} {\bibfield
  {journal} {\bibinfo  {journal} {Science}\ }\textbf {\bibinfo {volume}
  {341}},\ \bibinfo {pages} {1241214} (\bibinfo {year} {2013})}\BibitemShut
  {NoStop}%
\bibitem [{\citenamefont {Yolken}, \citenamefont {Dickerson},\ and\
  \citenamefont {Fuller~Torrey}(2009)}]{yolken2009toxoplasma}%
  \BibitemOpen
  \bibfield  {author} {\bibinfo {author} {\bibfnamefont {R.}~\bibnamefont
  {Yolken}}, \bibinfo {author} {\bibfnamefont {F.}~\bibnamefont {Dickerson}}, \
  and\ \bibinfo {author} {\bibfnamefont {E.}~\bibnamefont {Fuller~Torrey}},\
  }\bibfield  {title} {\enquote {\bibinfo {title} {Toxoplasma and
  schizophrenia},}\ }\href {\doibase 10.1111/j.1365-3024.2009.01131.x}
  {\bibfield  {journal} {\bibinfo  {journal} {Parasite Immunology}\ }\textbf
  {\bibinfo {volume} {31}},\ \bibinfo {pages} {706 -- 715} (\bibinfo {year}
  {2009})}\BibitemShut {NoStop}%
\bibitem [{\citenamefont {Forbes}\ \emph {et~al.}(1994)\citenamefont {Forbes},
  \citenamefont {Glaser}, \citenamefont {Cullen}, \citenamefont {Collins},
  \citenamefont {Warren}, \citenamefont {Christiansen},\ and\ \citenamefont
  {Marshall}}]{forbes1994duodenal}%
  \BibitemOpen
  \bibfield  {author} {\bibinfo {author} {\bibfnamefont {G.}~\bibnamefont
  {Forbes}}, \bibinfo {author} {\bibfnamefont {M.}~\bibnamefont {Glaser}},
  \bibinfo {author} {\bibfnamefont {D.}~\bibnamefont {Cullen}}, \bibinfo
  {author} {\bibfnamefont {B.}~\bibnamefont {Collins}}, \bibinfo {author}
  {\bibfnamefont {J.}~\bibnamefont {Warren}}, \bibinfo {author} {\bibfnamefont
  {K.}~\bibnamefont {Christiansen}}, \ and\ \bibinfo {author} {\bibfnamefont
  {B.}~\bibnamefont {Marshall}},\ }\bibfield  {title} {\enquote {\bibinfo
  {title} {Duodenal ulcer treated with {Helicobacter} pylori eradication:
  seven-year follow-up},}\ }\href {\doibase 10.1016/S0140-6736(94)91111-8}
  {\bibfield  {journal} {\bibinfo  {journal} {The Lancet}\ }\textbf {\bibinfo
  {volume} {343}},\ \bibinfo {pages} {258 -- 260} (\bibinfo {year}
  {1994})}\BibitemShut {NoStop}%
\bibitem [{IBG()}]{IBGE}%
  \BibitemOpen
  \href@noop {} {\enquote {\bibinfo {title} {{The Brazilian Institute of
  Geography and Statistics --- IBGE (Portuguese: Instituto Brasileiro de
  Geografia e Estatística)}},}\ }\bibinfo {howpublished} {{Available:}
  \url{https://www.ibge.gov.br/}},\ \bibinfo {note} {{Accessed:} 13 Jan
  2018}\BibitemShut {NoStop}%
\end{thebibliography}%

\clearpage
\clearpage
\linespread{1.2}
\setcounter{page}{1}
\setcounter{figure}{0}
\makeatletter
\renewcommand{\thefigure}{S\@arabic\c@figure}
\renewcommand{\thetable}{S\@arabic\c@table}

\onecolumngrid
\begin{center}
\Large{Supplementary Material for}\\
\vskip1pc
\large{\bf The hidden traits of endemic illiteracy in cities}\\
\vskip1pc
\normalsize{Luiz G. A. Alves, Jos\'e S. Andrade Jr., Quentin S. Hanley, Haroldo V. Ribeiro}
\end{center}

\begin{figure}[!ht]
\centering
\includegraphics[width=1\linewidth]{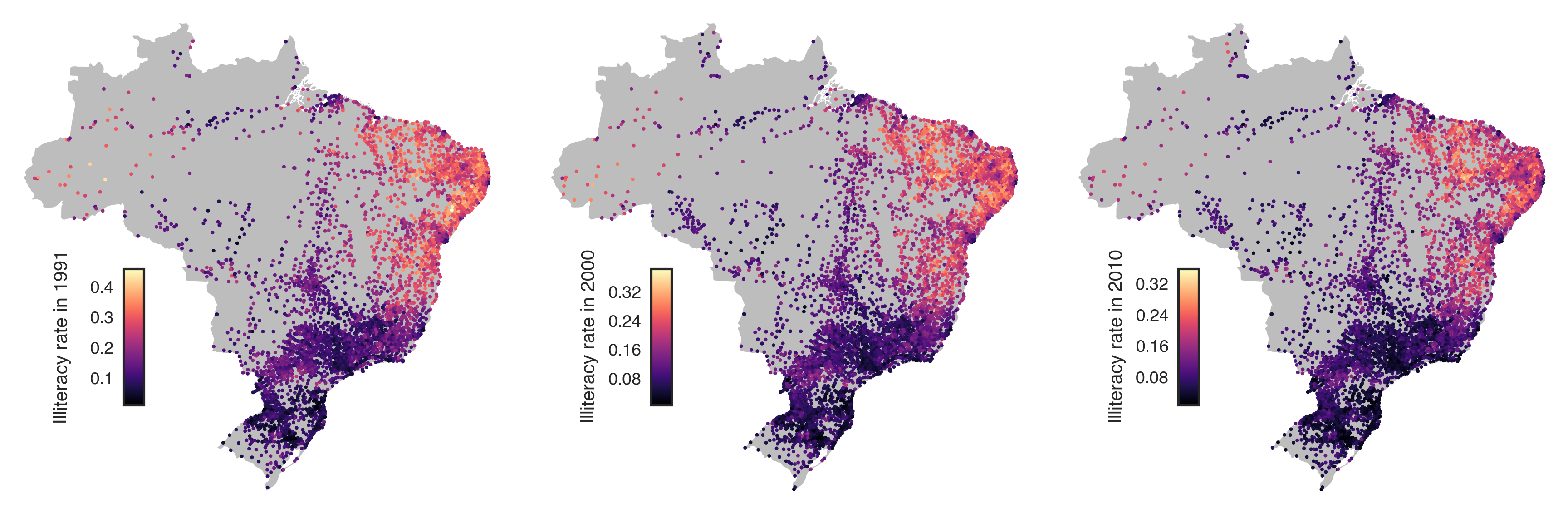}
\caption{{\bf Mapping illiteracy among Brazilian cities.} Each dot on these maps represents the location of a Brazilian city and the color code indicates the illiteracy rate at the place for the particular census year.}
\label{sfig1}
\end{figure}

\begin{figure}[!ht]
\centering
\includegraphics[width=1\linewidth]{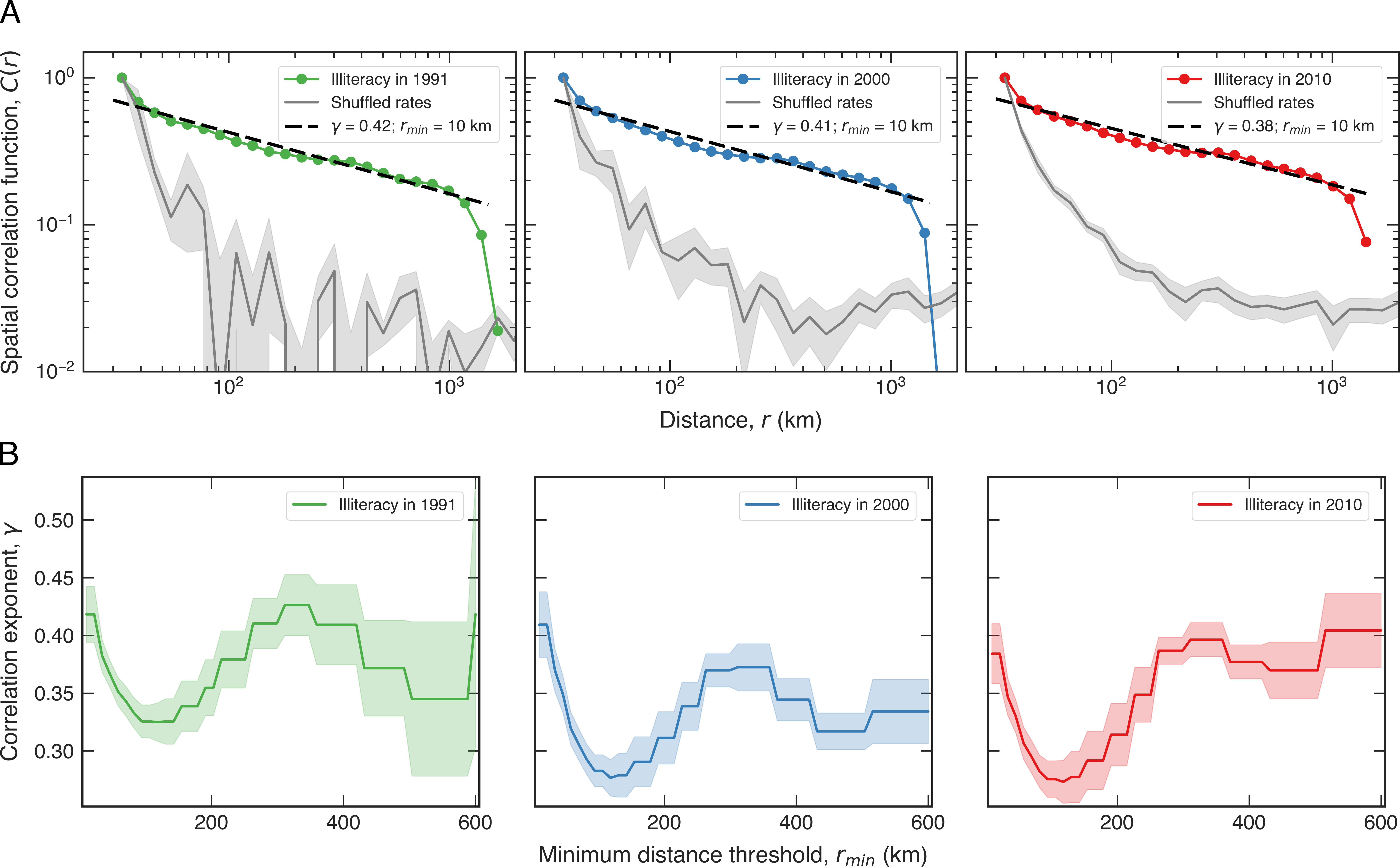}
\caption{{\bf Illiteracy rates are spatially long-range correlated.} (A) The correlation function $C(r)$ of the illiteracy rates in Brazilian cities for the years 1991, 2000, and 2010. The colorful dots are the empirical values of $C(r)$ and the dashed lines are a power-law decaying function, $C(r)\sim r^{-\gamma}$, adjusted to each data (via ordinary-least-square fits of the relationship $\ln C(r)$ versus $\ln r$) considering the range $10\leq r\leq1000~\text{km}$ (the values of $\gamma$ are shown in the plots). The gray curves represent the average values of $C(r)$ after random shuffling the rates among the cities (1000 realizations) and the shaded area stands for the 95\% confidence region. (B) The dependence of the values of $\gamma$ on the interval $r_{min}\leq r\leq1000~\text{km}$ employed to fit the relationship between $\ln C(r)$ and $\ln r$. The shared areas stand for 95\% bootstrap confidence intervals.}
\label{sfig2}
\end{figure}

\begin{figure}[!ht]
\centering
\includegraphics[width=1\linewidth]{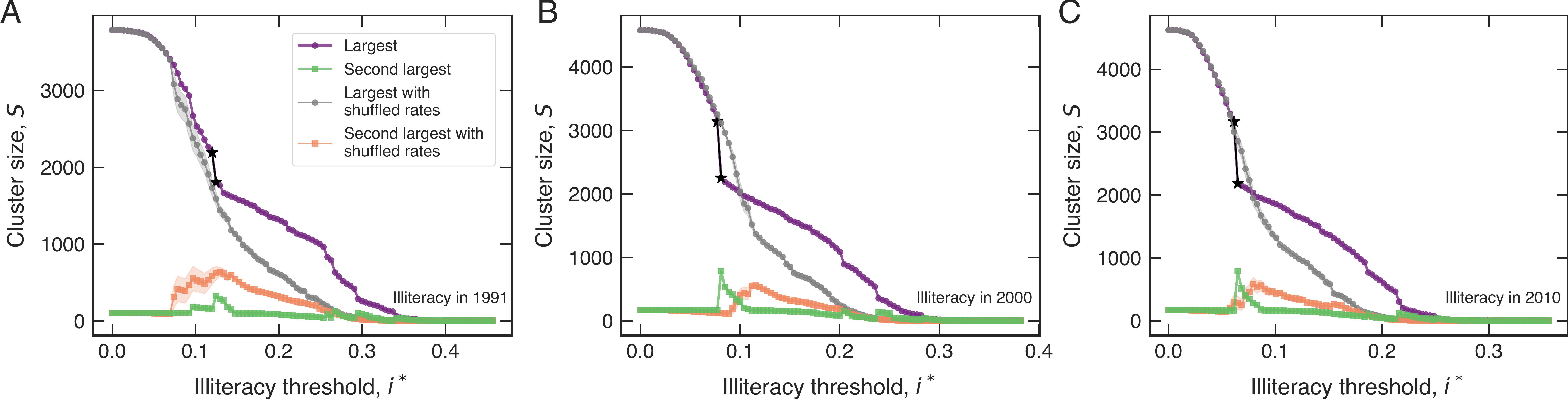}
\caption{{\bf Percolation-like transitions in illiteracy rates.} The purple curves in panels (A), (B), and (C) show the size of the largest cluster ($S$) as a function of the illiteracy threshold ($i^*$) for the three census years. The green curves represent the same for the second largest component. The transitions are indicated by star markers. The gray curves illustrate the behavior of the size of the largest cluster after shuffling the rates among cities. The orange curves represent the same for the second largest cluster. All results were obtained with $\varepsilon=48$~km.}
\label{sfig3}
\end{figure}

\begin{figure}[!ht]
\centering
\includegraphics[width=0.8\linewidth]{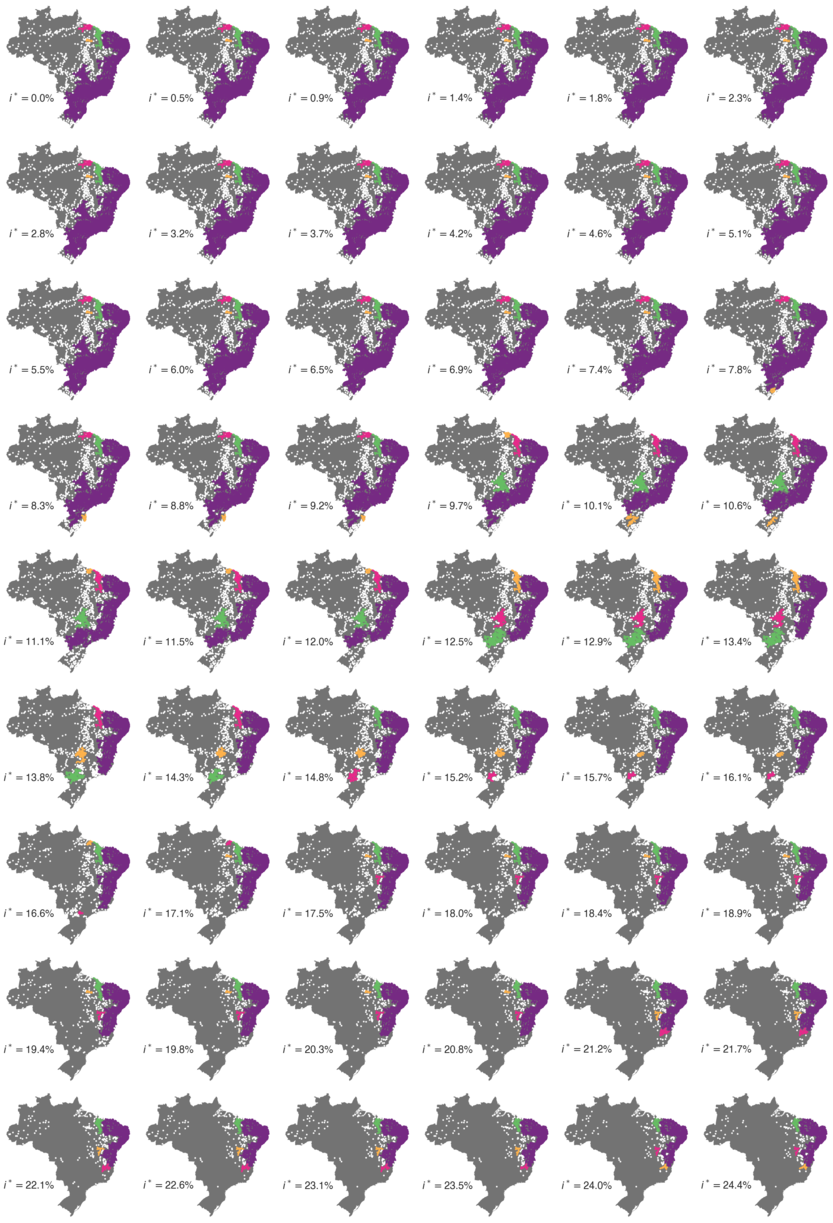}
\caption{{\bf The spatial clusters of cities having illiteracy rates larger than $i^*$.} The maps show the four largest identified clusters (colored in purple, green, pink, and yellow) for several values of $i^*$ (indicated by the plot). The white dots represent the cities in other smaller clustering components. All results are based on 1991 data with $\varepsilon=48$~km.}
\label{sfig4}
\end{figure}

\begin{figure}[!ht]
\centering
\includegraphics[width=0.8\linewidth]{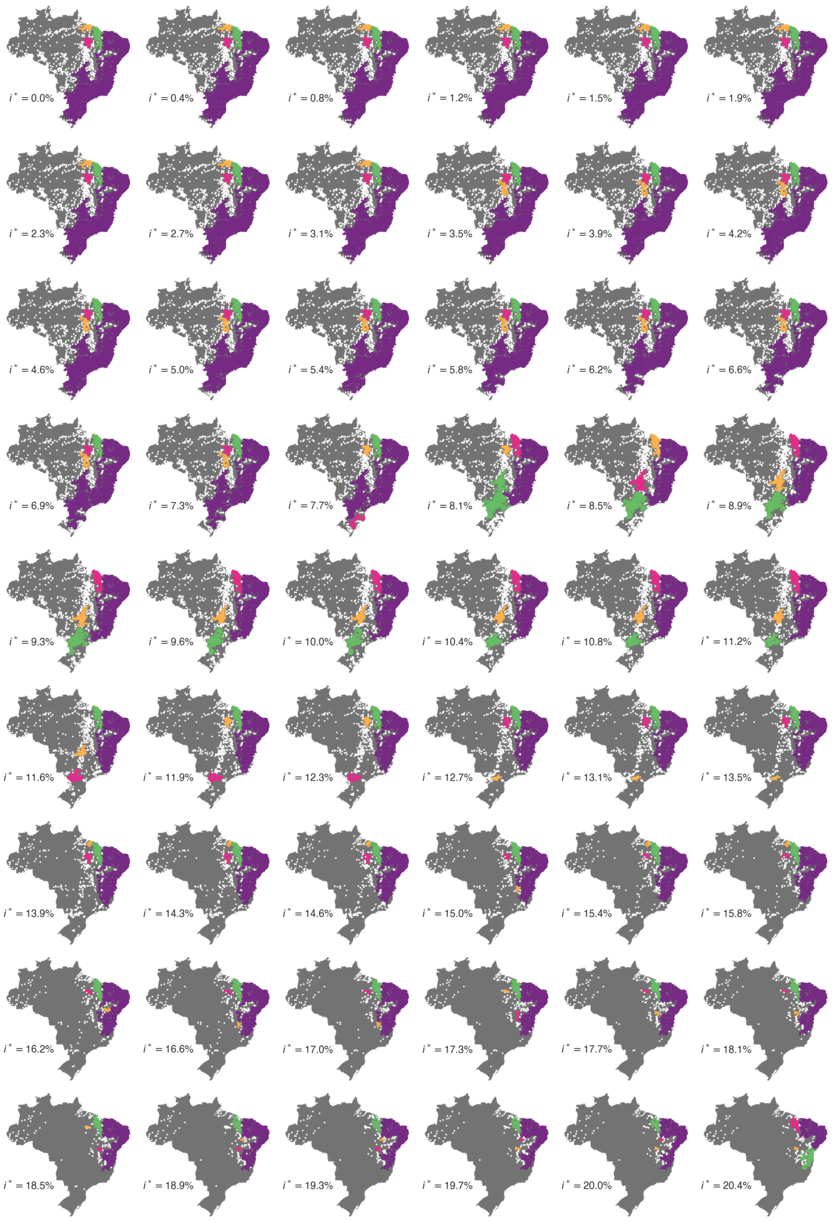}
\caption{{\bf The spatial clusters of cities having illiteracy rates larger than $i^*$.} The maps show the four largest identified clusters (colored in purple, green, pink, and yellow) for several values of $i^*$ (indicated by the plot). The white dots represent the cities in other smaller clustering components. All results are based on 2000 data with $\varepsilon=48$~km.}
\label{sfig5}
\end{figure}

\begin{figure}[!ht]
\centering
\includegraphics[width=0.8\linewidth]{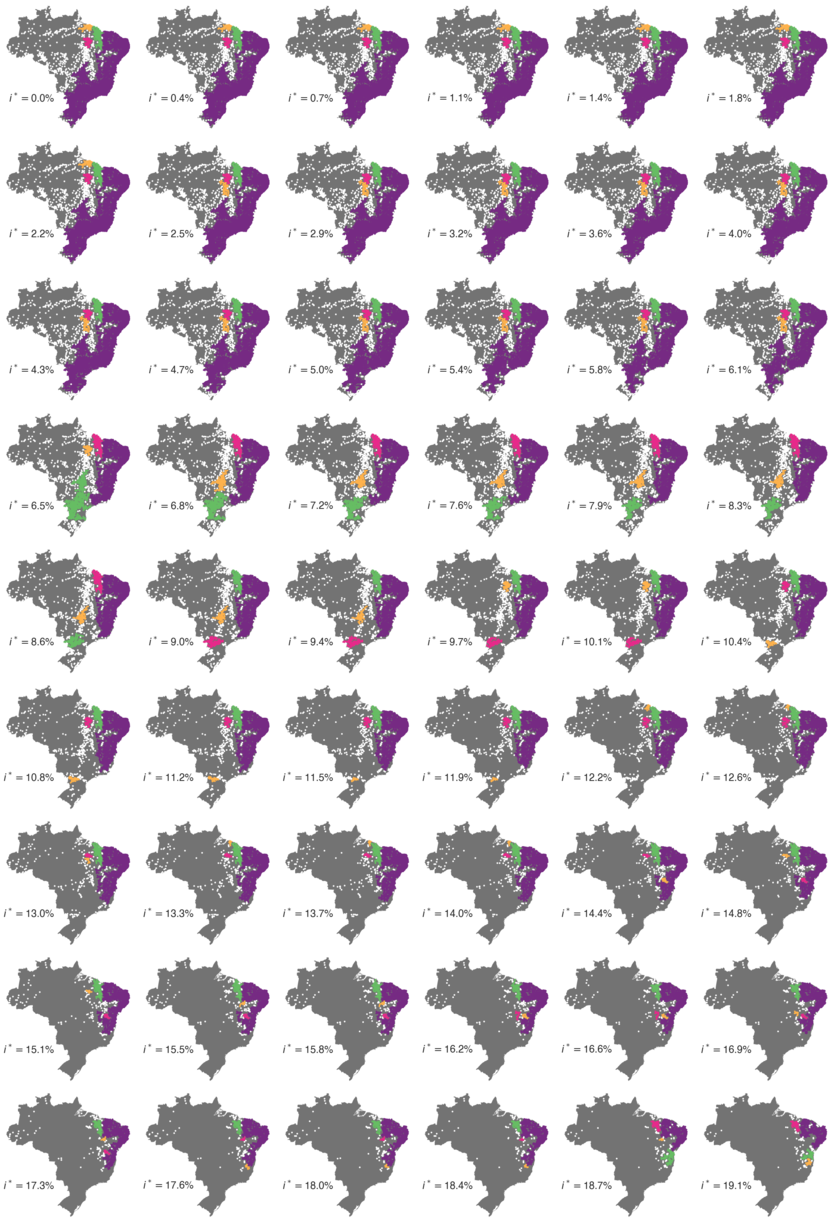}
\caption{{\bf The spatial clusters of cities having illiteracy rates larger than $i^*$.} The maps show the four largest identified clusters (colored in purple, green, pink, and yellow) for several values of $i^*$ (indicated by the plot). The white dots represent the cities in other smaller clustering components. All results are based on 2010 data with $\varepsilon=48$~km.}
\label{sfig6}
\end{figure}

\begin{figure}[!ht]
\centering
\includegraphics[width=0.8\linewidth]{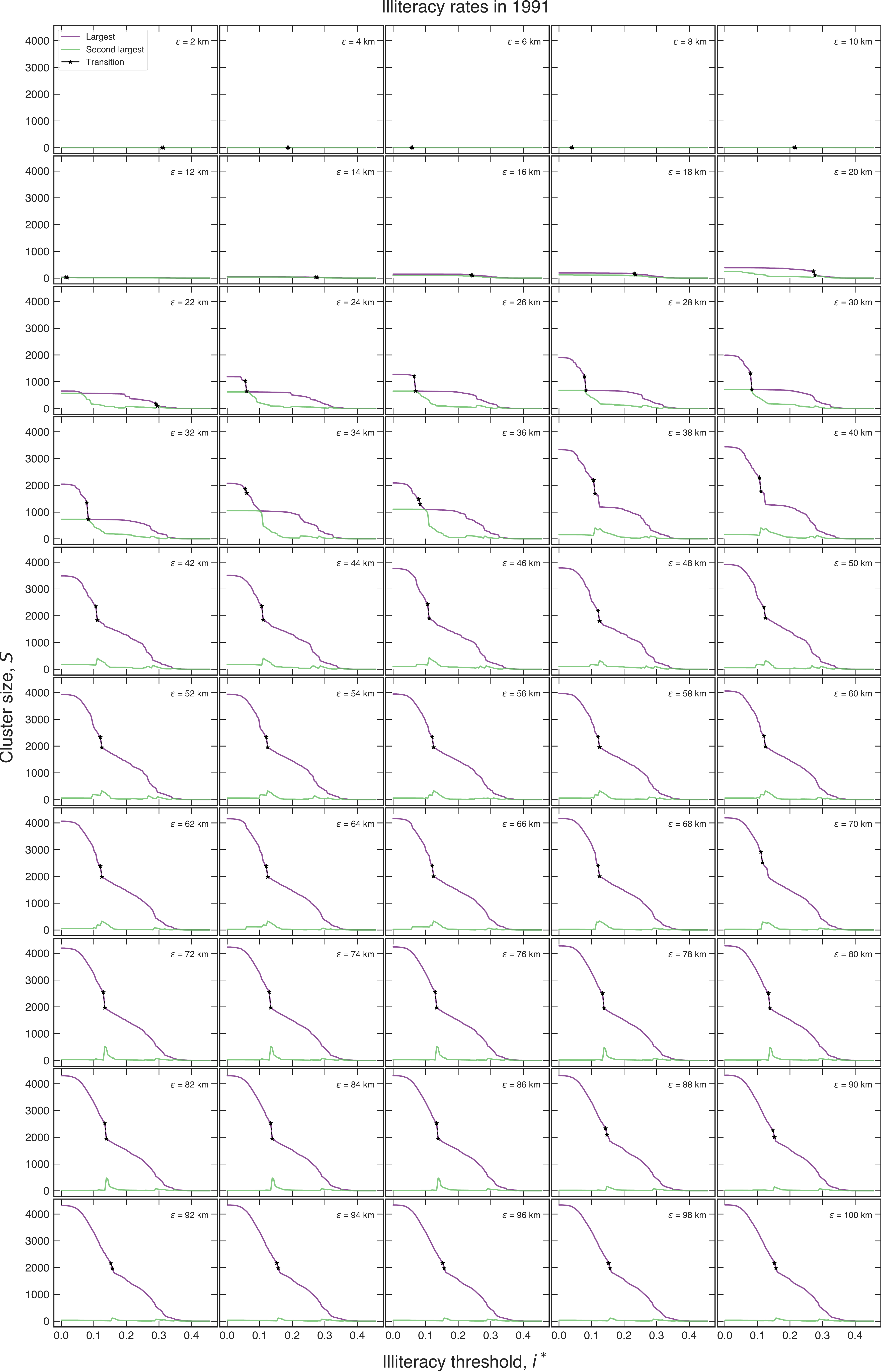}
\caption{{\bf Percolation-like transitions in illiteracy rates: changes with the DBSCAN parameter $\varepsilon$.} The purple (green) curves show the dependence of the size of the (second) largest cluster on the illiteracy threshold $i^*$ for different values of the parameter $\varepsilon$ (as indicated in the plots). These results are based on 1991 data.}
\label{sfig7}
\end{figure}

\begin{figure}[!ht]
\centering
\includegraphics[width=0.8\linewidth]{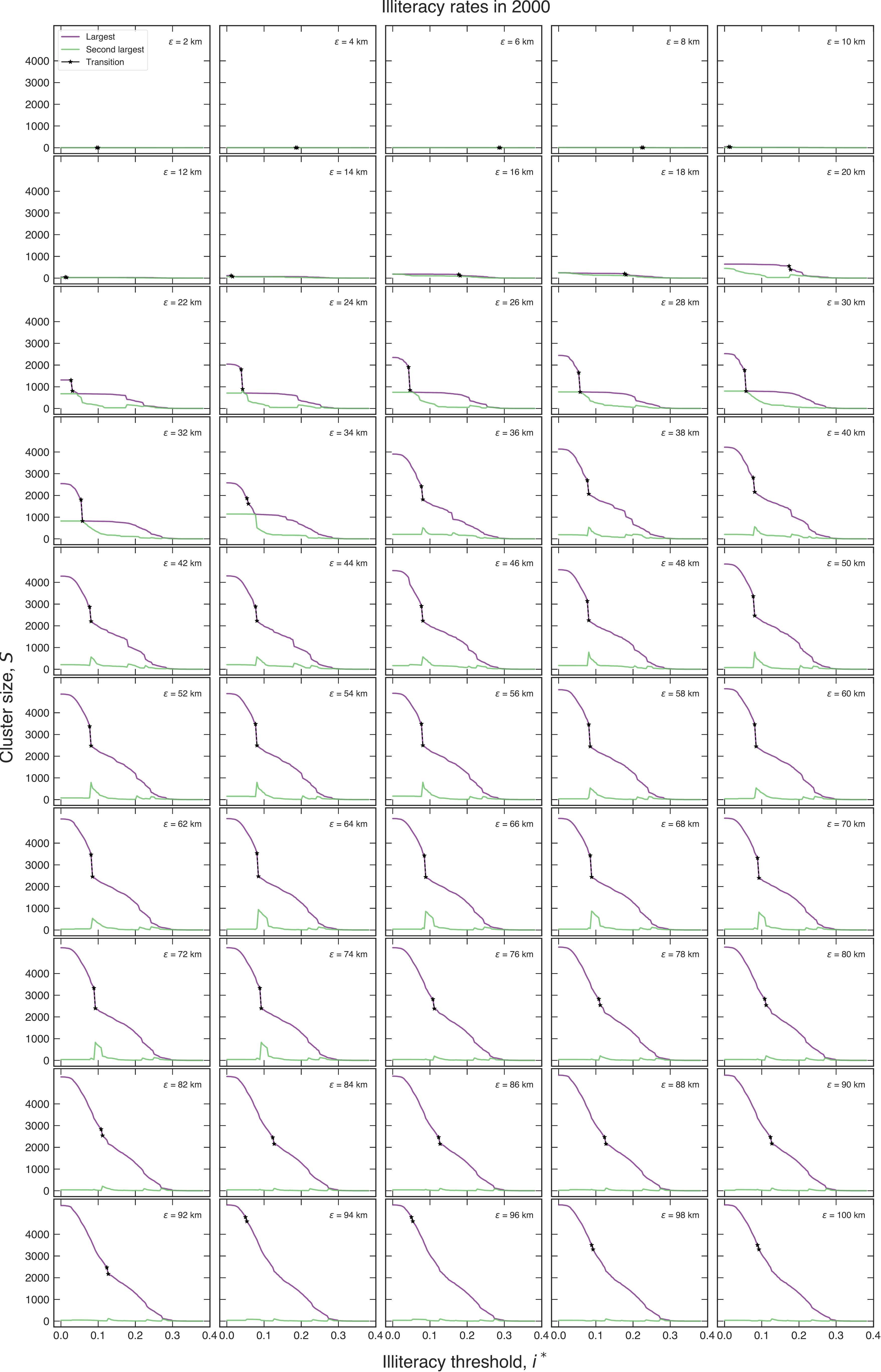}
\caption{{\bf Percolation-like transitions in illiteracy rates: changes with the DBSCAN parameter $\varepsilon$.} The purple (green) curves show the dependence of the size of the (second) largest cluster on the illiteracy threshold $i^*$ for different values of the parameter $\varepsilon$ (as indicated in the plots). These results are based on 2000 data.}
\label{sfig8}
\end{figure}

\begin{figure}[!ht]
\centering
\includegraphics[width=0.8\linewidth]{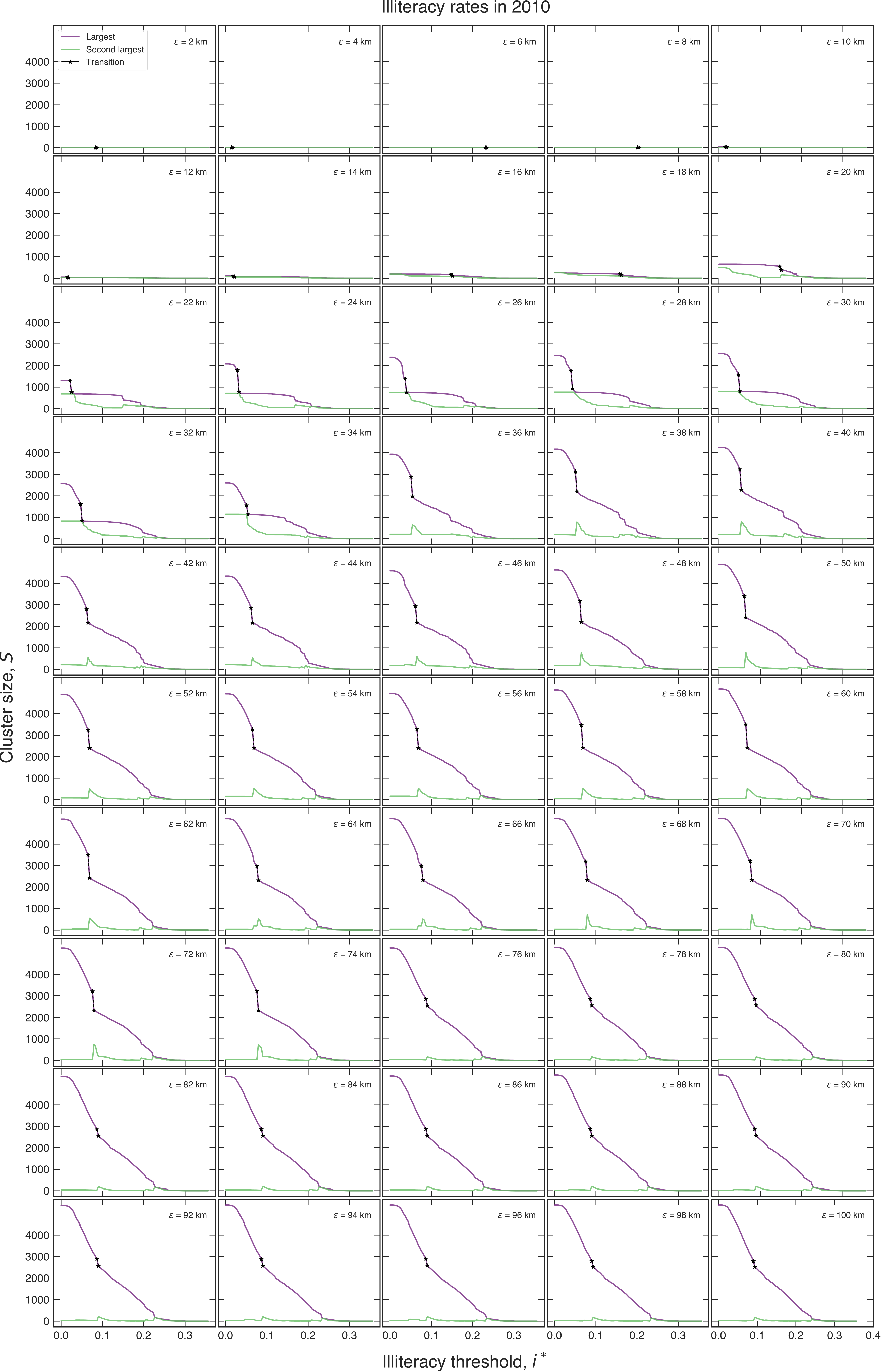}
\caption{{\bf Percolation-like transitions in illiteracy rates: changes with the DBSCAN parameter $\varepsilon$.} The purple (green) curves show the dependence of the size of the (second) largest cluster on the illiteracy threshold $i^*$ for different values of the parameter $\varepsilon$ (as indicated in the plots). These results are based on 2010 data.}
\label{sfig9}
\end{figure}

\begin{figure}[!ht]
\centering
\includegraphics[width=0.8\linewidth]{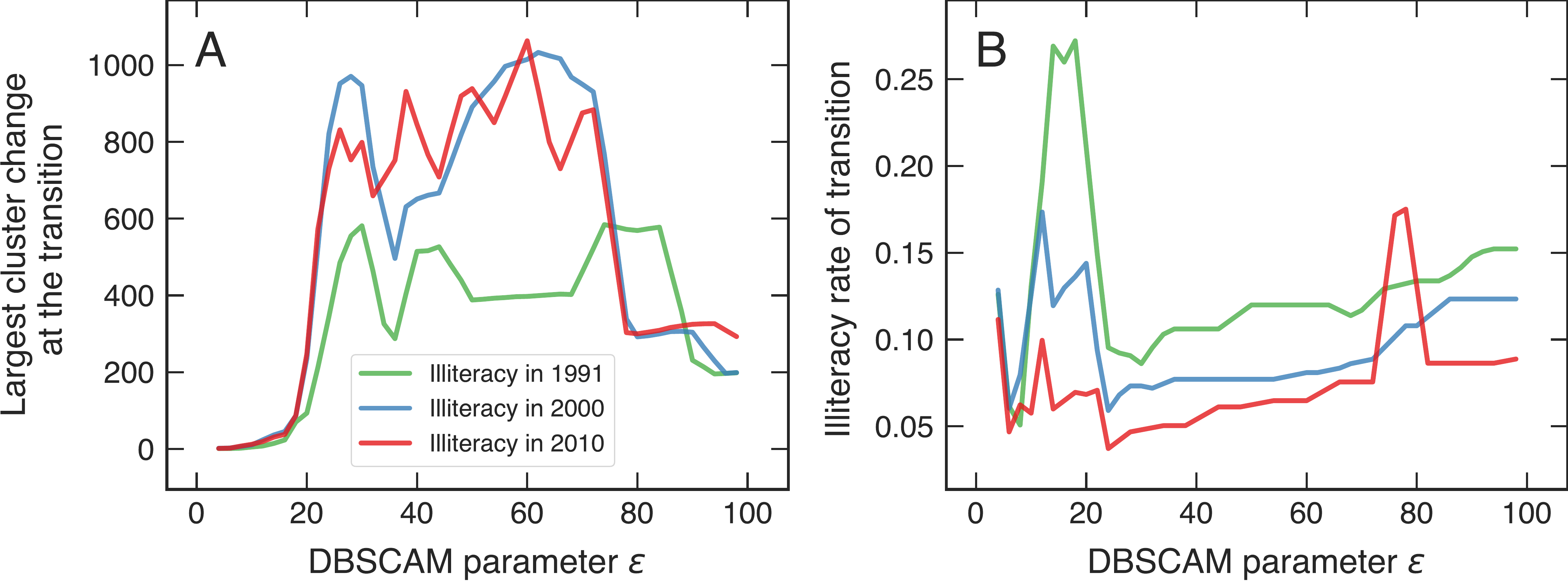}
\caption{{\bf Percolation-like transitions in illiteracy rates: changes with the DBSCAN parameter $\varepsilon$.} Panel (A) shows the dependence of the jump in the size of the largest cluster near the transition as a function of the parameter $\varepsilon$ for the three census years. Panel (B) shows the dependence of the illiteracy threshold $i^*$ in which the transition took place as a function of the parameter $\varepsilon$ for the three census years. We notice that the transitions are very similar when $25<\varepsilon<75$~km.}
\label{sfig10}
\end{figure}

\begin{figure}[!ht]
\centering
\includegraphics[width=0.8\linewidth]{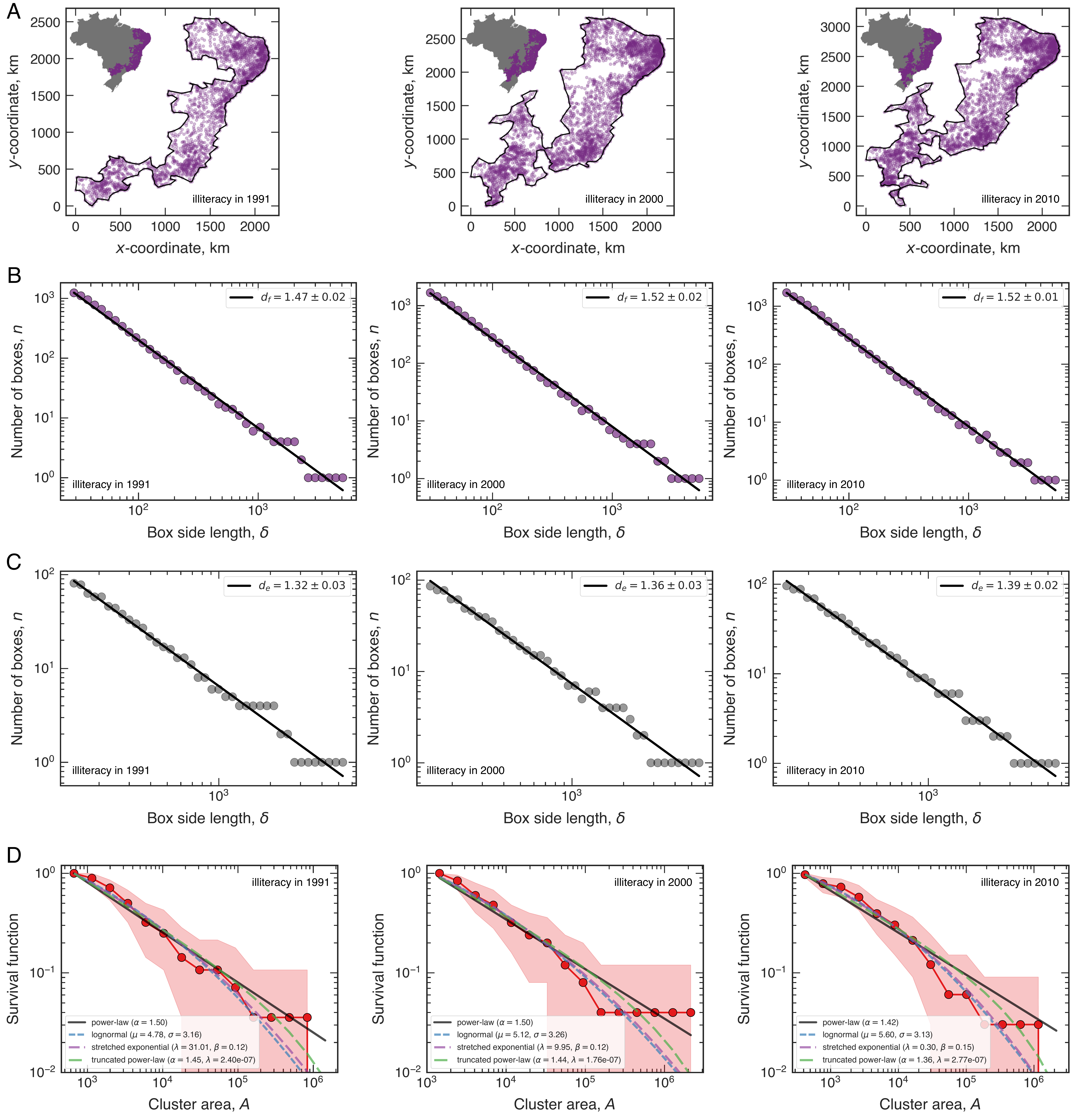}
\caption{{\bf The critical exponents of illiteracy clusters: results for each census year.} Panel (A) shows the shapes of the largest clusters immediately before the transition for each year, where purple dots represent cities within the largest component, and the black line is the concave hull enveloping the largest cluster. Panel (B) shows the relationships between the number of boxes $n$ necessary to cover the largest cluster as a function of the box side length $\delta$ for each year. The continuous lines are power-law fits ($n\sim\delta^{-d_f}$), where the value of $d_f$ (indicated in the plots) represents the box-counting fractal dimension. Panel (C) shows the analogous of the previous panel when considering only the concave hull points. The continuous lines are power-law fits ($n\sim\delta^{-d_e}$), where the value of $d_e$ (indicated in the plots) represents the box-counting fractal dimension of the hull points. Panel (D) shows the survival functions (complementary cumulative distribution) of the area of clusters ($A$) near the criticality for each census year. The different lines represent four probability distributions adjusted to the empirical data (names and parameters are shown within the plot legend). The shaded areas stand for the 95\% bootstrap confidence region of the survival functions.}
\label{sfig11}
\end{figure}

\end{document}